\documentclass[%
 aip,
 amsmath,amssymb,
 reprint,%
]{revtex4-1}

\usepackage{graphicx}
\usepackage{dcolumn}
\usepackage{bm}

\usepackage[utf8]{inputenc}
\usepackage[T1]{fontenc}
\usepackage{mathptmx}
\usepackage{etoolbox}
\usepackage{color}
\usepackage[table]{xcolor}
\usepackage{booktabs}

 \usepackage{soul}

\newcommand{\Up}{\mathbf{U}}
\newcommand{\Dn}{\mathbf{D}}

\makeatletter
\def\@email#1#2{%
 \endgroup
 \patchcmd{\titleblock@produce}
  {\frontmatter@RRAPformat}
  {\frontmatter@RRAPformat{\produce@RRAP{*#1\href{mailto:#2}{#2}}}\frontmatter@RRAPformat}
  {}{}
}%
\makeatother
\begin{document}

\preprint{AIP/123-QED}

\title[Glassy Landscape Maps]{Mapping out the glassy landscape of a mesoscopic elastoplastic model}
\author{D. Kumar}
\email{dheeraj.kumar@espci.fr}
\author{S. Patinet}%
\affiliation{PMMH, CNRS, ESPCI Paris, Universit\'e PSL, Sorbonne Universit\'e, Universit\'e Paris Cit\'e}%

\author{C. E. Maloney}
\affiliation{NorthEastern University, Boston, Massachusetts 02115, USA}

\author{I. Regev}
\affiliation{Department of Solar Energy and Environmental Physics, Jacob Blaustein Institutes for Desert Research, Ben-Gurion University of the Negev,
Sede Boqer Campus 84990, Israel}%
\author{D. Vandembroucq}%
\affiliation{PMMH, CNRS, ESPCI Paris, Universit\'e PSL, Sorbonne Universit\'e, Universit\'e Paris Cit\'e}%

\author{M. Mungan}
\affiliation{Institut f\"{u}r angewandte Mathematik, Universit\"{a}t Bonn, Endenicher Allee 60, 53115 Bonn, Germany
}

\date{\today}

\begin{abstract}
We develop a mesoscopic model to study the plastic behavior
of an amorphous material under cyclic loading. The model is depinning-like and driven by 
a disordered thresholds dynamics which are coupled by long-range elastic interactions. 
We propose a simple protocol of ``glass preparation'' which allows us to mimic thermalisation at high temperature, as well as aging at vanishing temperature. Various levels of glass stabilities (from brittle to ductile) can be achieved by tuning the aging duration. The aged 
glasses are then immersed into a quenched disorder landscape and serve as initial configurations for various protocols of mechanical loading by shearing.
The dependence of the plastic behavior upon monotonous loading is recovered. The behavior under cyclic loading is studied for different ages and system sizes. The size and age dependence of the irreversibility transition is discussed. A thorough characterization of the disorder-landscape is achieved through the analysis of the transition graphs, which describe the plastic deformation pathways under athermal quasi-static shear. In particular, the analysis of the stability ranges of the strongly connected components of the transition graphs reveals the emergence of a phase-separation like process associated with the aging of the glass. Increasing the age and hence  stability of  the initial glass, results in a gradual break-up of the  landscape of dynamically accessible stable states
into three distinct regions: one region centered around the initially prepared glass phase, and two additional regions, characterized by well-separated ranges of positive and negative plastic strains, each of which is accessible only from the initial glass phase by passing through the stress peak in the forward, respectively, backward shearing directions.

\end{abstract}

\maketitle

\section{\label{sec:intro}Introduction}
Understanding the response of a disordered solid to an externally imposed forcing,
such as stress or strain, is important in order to characterize the transitions between rigid and flowing states in a
wide variety of soft matter systems. Examples for such
behavior include the jamming transition in granular materials~\cite{behringer2018physics}, the yielding transition in amorphous solids~\cite{bonn2017yield,Nicolas-RMP18}, and the depinning transition of a pinned elastic interface, such as flux-lines in type II superconductors~\cite{Blatter}. 

The interplay between the deformation energy cost and gain, as the disordered solid adapts to the imposed forcing by deforming, gives rise to rich dynamics on a complex energy landscape. For small loading, the response of the solid is largely elastic, characterized by few plastic deformation events. However as the loading is increased, plastic deformations start to proliferate and  eventually this leads to yielding and flow. The manner in which the transition to yielding occurs has been found to depend strongly on the degree of initial annealing, i.e. aging, of the sample~\cite{VR-PRB11,ozawa2018random, Wyart-PRE18, Fielding2020, yeh2020glass, bhaumik2021role}.

Over the last years a large body of experimental~\cite{keim2013yielding,keim2014mechanical,knowlton2014microscopic,nagamanasa2014experimental,keim2018return} and numerical~\cite{Shi-Falk-PRL05,shi2007evaluation,fiocco2013oscillatory,regev2013onset,wisitsorasak2012strength,jaiswal2016mechanical,kawasaki2016macroscopic,regev2015reversibility,jin2018stability,procaccia2017mechanical,leishangthem2017yielding,ozawa2018random,parmar2019strain,barbot2020rejuvenation,bhaumik2021role,yeh2020glass,bhaumik2022avalanches,bhaumik2022yielding}
work has been carried out to understand the nature of the yielding transition in amorphous solids. These results reveal an intriguingly complex and dynamical spectrum of response that,  besides its dependence on the degree of annealing and the amount of loading, also shows dependence on history, as well as system size and dimensionality. 

Of special recent interest has been the response of amorphous solids to cyclic shear, in particular under athermal quasistatic (AQS) conditions~\cite{fiocco2013oscillatory,regev2013onset,PRIEZJEV2013,regev2015reversibility,leishangthem2017yielding,parmar2019strain,yeh2020glass,bhaumik2021role}. 
Experiments and simulations show~\cite{pine2005chaos, corte2008random,Keimetal2011,regev2013onset,fiocco2013oscillatory,PRIEZJEV2013,lavrentovich2017period} that for small oscillatory strain amplitudes, the solid settles into a cyclic response after just a few driving cycles. As the strain amplitude is increased, the transients to cyclic response become increasingly longer, and multi-periodic response, i.e. cycles that repeat every $T>1$ driving periods, starts to emerge. This behavior continues until a critical strain amplitude is reached, beyond which cyclic response is no longer attainable and particles start to diffuse across the sample. The transition from cyclic to diffusive behavior has been found to be rather sharp and is called the {\em irreversibility transition}. 

At the same time, a cyclic response to periodic loading can also be regarded as a form of memory which encodes information about the forcing that produced the response~\cite{keim2019memory}. Such memory effects have been observed experimentally as well as numerically, in periodically sheared amorphous solids, colloidal suspensions as well as other soft-condensed matter systems~\cite{pine2005chaos,corte2008random,Keimetal2011,Fiocco2014,Mungan-PRL19,keim2018return,brown2019reversible,Regev2021}.

Along with atomistic models of amorphous solids, spatially coarse-grained {\it mesoscopic} elastoplastic models~\cite{Nicolas-RMP18} have been introduced. 
Due to their conceptual simplicity, mesoscopic models are appealing both from a numerical as well as a theoretical perspective. 
Initially, the main goal of these models has been to capture the response under monotonous loading by shear strain. However more recently, mesoscopic models have been constructed that study the response under oscillatory shear~\cite{chenliu2020,Maloney2021}. 
In order to be able to realistically capture cyclic response, particularly key features of the irreversibility transition, a prescription for replacing mesoscopic elements once they have yielded has to be provided. We will refer to the available choices generated by a given replacement prescription as the ``landscape'' of the mesoscopic model. 

Of special interest are two recently introduced elastoplastic models that study the response under cyclic shear: The  model by Liu {\em et al.}~\cite{chenliu2020} assumes that mesoscopic elements that yielded are replaced at random and hence irreversibly, while the model of Khirallah {\em et al.}~\cite{Maloney2021} is complementary in that it is fully deterministic: elements that yielded are replaced by ones with identical, i.e. non-random, yield stresses.
The only source of randomness being the initial internal stress configuration. Thus in terms of the landscape terminology the model of Liu {\em et al.} has a totally random disorder-landscape, while the one of Khirallah {\em et al.} is totally ordered.
Despite of these differences, 
both models nevertheless recover key features of the response of amorphous solids to cyclic shear, such as the irreversibility transition and divergence of lengths of transients as the transition is approached.   
Let us finally note that these types of mesoscopic models have been used as a starting point for developing even further coarse-grained models, such as the recently introduced stochastic mesostate models~\cite{sastry2020mesoland,mungan2021metastability,parley2021}. These models, too, capture key features of the irreversibility transition of amorphous solids under oscillatory shear.  

Here we present a depinning-like mesoscopic elastoplastic model with a quenched disorder landscape. Our model therefore interpolates between the two types of landscapes considered before. Specifically, the  model we consider has two features : (i) a local yielding protocol which allows us to mimic thermalization and aging, and thereby to tune the history of our samples, (ii) the quenched disordered landscape, which allows us to capture in rather great detail the transients and the evolution to cyclic response in terms of the localized plastic events. 

As in previous work\cite{chenliu2020,Maloney2021}, we first focus on the stress response under monotonous loading by an externally applied shear strain. Our model recovers the {\em brittle-to-ductile} transition: as our initial glass is increasingly aged better, the stress response exhibits a stress peak that gets more pronounced with the duration of aging.

We next focus on the irreversibility transition under oscillatory shear and its dependence on both the degree of annealing and system size. We find that for poorly- and moderately-aged samples, the transient times to cyclic response diverge as the irreversibility transition is approached. In the case of poorly-aged samples, this divergence follows a power-law with an 
exponent that is comparable with estimates obtained in recent works~\cite{regev2013onset,regev2015reversibility,kawasaki2016macroscopic,Maloney2021}.

We finally turn to a more detailed comparison between the disorder landscape of mesoscopic and atomistic models. To this end we make use of the fact that the AQS dynamics of driven disordered systems has a natural representation in terms of a transition graph, the $t$-graph~\cite{munganterzi2018structure, MunganWitten-PRE19, Mungan-PRL19, Regev2021}. 
The AQS dynamics is thereby encoded into the topology of the $t$-graph and 
provides a unified setting within which we can compare in great detail the properties of the disorder landscapes underlying our mesoscopic and atomistic models. 

We perform such comparisons by focusing on a particular topological feature of the the $t$-graph, its {\em strongly connected components (SCCs)}. An SCC is a collection of mechanically stable configurations, actually elastic branches, 
which are connected in a bi-directional manner by plastic deformation pathways: a pair of configurations belongs to the same SCC, if there is a deformation pathway that leads from one to the other and back. Hence the plastic events triggered by transitions between states belonging to the same SCC are mechanically reversible, while transitions connecting different SCCs are irreversible~\cite{Regev2021}. Any periodic response must necessarily be confined to a single SCC, and therefore the size of the SCCs and their dynamic accessibility is a limiting factor for the length of transients and the cyclic strain amplitudes at which cyclic behavior can be attained~\cite{Regev2021}. 

The manuscript is organized as follows. In section~\ref{sec:depinning-model}, we present the mesoscopic model developed for the present study. In section~\ref{sec:glass-preparation}, we detail the protocols of preparation that allow us to mimic annealing at high temperature and aging at vanishing temperature, respectively. In section~\ref{sec:monotonous-loading}, we show that varying the level of aging allows us to recover upon monotonous loading either  a ductile response or a stress peak followed by a softening branch.
In section~\ref{sec:cylic-driving}, we focus on the analysis of the irreversibility transition upon cyclic driving. In particular we discuss the dependence of the transition on sample size and preparation by aging. Finally, we give in section~\ref{sec:t-graphs} a characterization of the limit cycles observed  based on the transition graphs, as recently proposed in the context of atomistic simulations~\cite{Mungan-PRL19}. Let us stress that our main aim in presenting our mesoscopic model is not to {\em quantitatively} reproduce the results of atomistic simulations. Rather, our intention has been to use the molecular dynamic results and athermal quasi-static shear as a qualitative reference against which we compare the different results  obtained in the present work.

\section{A depinning-like model for amorphous plasticity\label{sec:depinning-model}}

We consider a scalar 2D lattice-based mesoscale elasto-plastic model. The physics of this class of models relies on the coupling between a threshold dynamics and an elastic interaction induced by the incremental local plastic slip which arises as a result of a mechanical instability~\cite{Nicolas-RMP18}.

\begin{figure}
\includegraphics[width=0.4\textwidth]{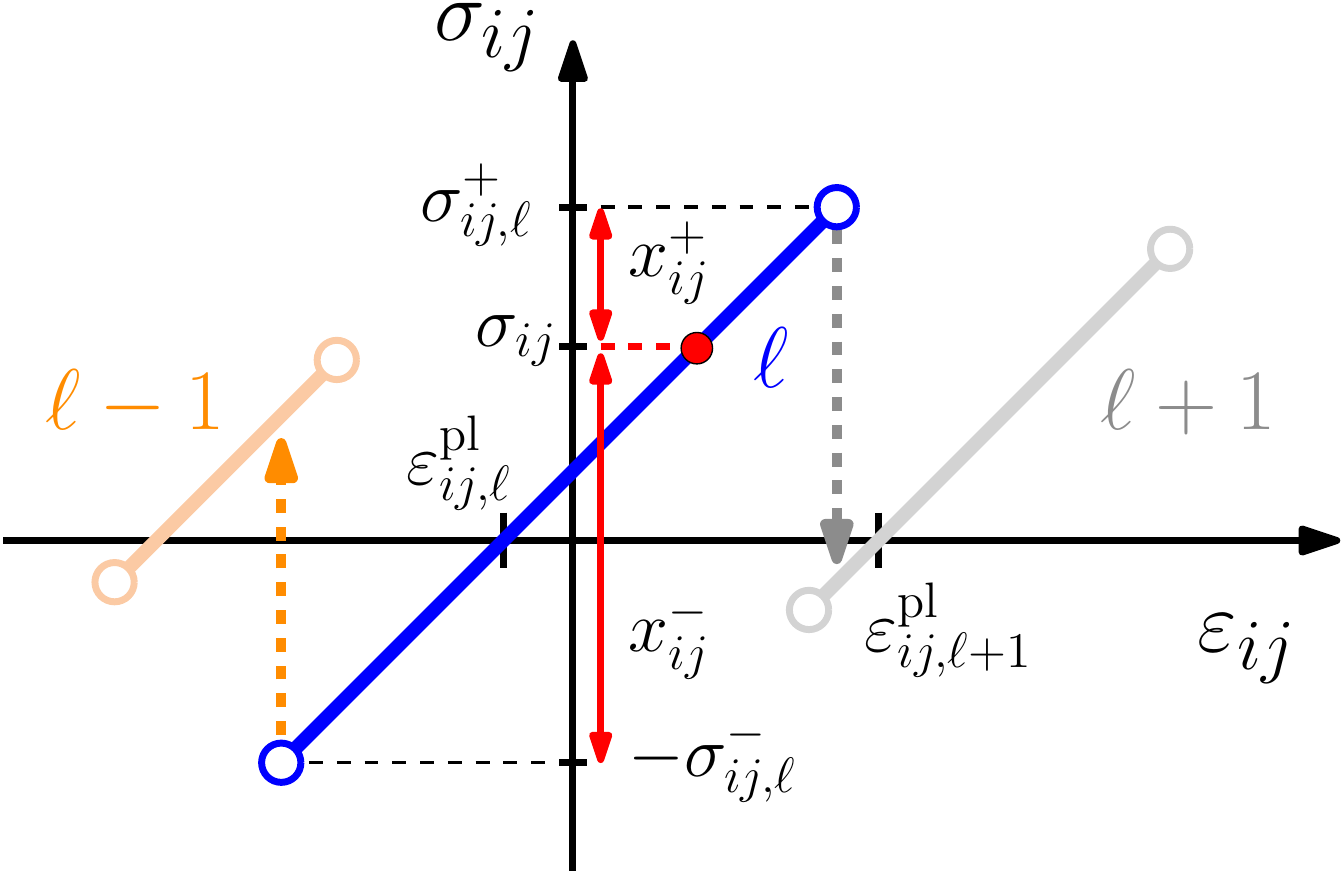}
\caption{Local elastic branches associated with a cell $(i,j)$. Each elastic branch $\ell$ is characterized by a pair of stress thresholds $\sigma^\pm_{ij,\ell}$ and a plastic strain $\epsilon^{\rm pl}_{ij,\ell}$, which prescribe the behavior of the local stress $\sigma_{ij}$ under elastic strain $\epsilon^{\rm el}_{ij}=\epsilon_{ij} - \epsilon^{\rm pl}_{ij}$, as shown for the branch labeled $\ell$ in the figure. When the stress reaches the upper or lower stress threshold, a transition to the corresponding neighbouring branches, $\ell \pm 1$ occurs. The current stress state of the cell is denoted by a red filled symbol on the elastic branch $\ell$. This allows us to define the local plastic strengths $x_{ij}^+=\sigma_{ij}^+ - \sigma_{ij}$ and  $x_{ij}^-=\sigma_{ij}^- + \sigma_{ij}$ which give the distance to threshold in the forward and backward directions, respectively. The slopes of the local branches are identical and equal to $2\mu$. }
\label{fig:sketch-elastic branch}
\end{figure}

More specifically, we consider a square grid of $N\times N$ cells of size $a\times a$. The model is scalar, so that we account for one and only one shear direction, along which we can shear the system forward and backwards. 
We assume a uniform shear modulus $\mu$. Each individual cell $(i,j)$ is characterized by a stack of local elastic branches indexed by a variable $\ell$, each of which relates the local stress $\sigma_{ij}$ to the local strain $\varepsilon_{ij}$, as shown in Fig.~\ref{fig:sketch-elastic branch}. The stability of each such local elastic branch $\ell$ is limited by two bounds: a maximum stress threshold $\sigma_{ij,
\ell}^+$, and a minimum stress threshold $-\sigma_{ij,\ell}^-$. Note that in order to ease notation, whenever no explicit reference to a particular branch number $\ell$ is made, we will omit it in the following. The two thresholds $\sigma_{ij}^+$ and $\sigma_{ij}^-$ are drawn from a random distribution with support in $\mathbf{R}^+$ so as to ensure $-\sigma_{ij}^- < \sigma_{ij}^+$, i.e. the existence of a stability domain for the cell $(i,j)$.

In the present model, the local stress $\sigma_{ij}$ experienced by the cell $(i,j)$ originates from two distinct contributions: a global stress $\Sigma$ due to the external loading, and an internal stress associated to the interactions with other cells, so that $\sigma_{ij} = \Sigma + \sigma_{ij}^{\rm int}$. The latter contribution fluctuates spatially and is by definition of zero average so that we have $\overline{\sigma_{ij}^{int}}=0$, and therefore $\overline{\sigma_{ij}} = \Sigma $. Here $\overline{A}$ denotes the spatial average of the observable $A$.

Due to the external loading and the stress interactions, the local stress $\sigma_{ij}$ is in general non-zero so that the amount of (external) stress that needs to be  applied in order to reach one of the boundaries of the elastic branch is not a priori equal to the stress thresholds $\sigma_{ij}^+, \sigma_{ij}^-$. Instead, it is given by the {\em local plastic strengths} in the positive and negative directions, which we define as 
$x^+_{ij} = \sigma_{ij}^+ - \sigma_{ij}$ and $x^-_{ij} = \sigma_{ij} + \sigma^-_{ij}$,  respectively. Note that for a mechanically stable configuration we require that $-\sigma^-_{ij} < \sigma_{ij} < \sigma^+_{ij}$, so that the quantities $x^\pm_{ij}$ must be positive in that case. 

The separation $\Delta \varepsilon$ 
between two neighboring local elastic branches that belong to a given cell $(i,j)$ defines the local plastic strain $\varepsilon^{\rm pl}_{ij, \ell}$ experienced by the cell after the local stress has reached threshold in one or the other direction.

{\it Stress interaction} \--- Local plastic strains are generated within an elastic matrix (the other cells of the lattice). This incompatibility induces an internal Eshelby stress field of quadrupolar symmetry~\cite{Eshelby57}. Since we assume homogeneous elasticity, the elastic response to a unit plastic slip can be computed once and for all. The internal stress thus directly arises from the convolution of the field of plastic strain with the Green function of Eshelby  stresses. The latter is computed from the discrete Fourier Transform of the analytical solution in the reciprocal space. Details on the implementation and a discussion can be found in Refs.~\cite{Talamali-CRM12,TPRV-PRE16}. 

The typical stress drop associated to a rearrangement of plastic strain $\Delta\varepsilon$ is of order $\mu \Delta \varepsilon$. 
For the sake of comparisons with atomistic simulations, we consider here $\mu=10$, a typical value observed in Lennard-Jones binary model glasses~\cite{regev2013onset,Patinet-PRL2020-Bauschinger}

{\it Random landscape} \--- The stress thresholds are drawn from a random distribution $P(\sigma^\pm)$. Here we consider a Weibull distribution of parameters $\lambda=1.0,  k=2.0$, where $\lambda$ and $k$ are constants in the cumulative density function given by $ 1 - e^{-(\sigma^{\pm}/\lambda)^{k}} $. 
The plastic strain increment $\Delta \varepsilon = \varepsilon^{\rm pl}_{ij,\ell + 1} - \varepsilon^{\rm pl}_{ij,\ell} $ between two neighbor elastic branches $\ell$ and $\ell+1$ is also a random variable, {\em cf.} Fig.~\ref{fig:sketch-elastic branch}. We choose it to be correlated to the two plastic thresholds associated with the transition $\ell \to \ell+1$, i.e. $\sigma_{ij,\ell}^+$ in the forward direction and $\sigma_{ij,\ell+1}^-$ in the backward direction. 
More specifically, we choose $\Delta\varepsilon$ from a uniform distribution in $[0,\Delta\varepsilon_{max}]$ with $\Delta\varepsilon_{max}=\eta (\sigma_{ij,\ell}^++\sigma_{ij,\ell+1}^-)/(2\mu)$, 
where 
$\eta$ is a tunable parameter.
Note that the parameter $\eta$ 
thus controls the strength of the elastic interaction~\cite{Talamali-CRM12,Budrikis-NatComm17}:
the larger $\eta$, the larger the short range stress kicks that trigger the avalanche, but also the larger the amplitude of the mechanical noise   arising from the small positive and negative  contributions of the long range  stress interaction. We have set $\eta = 1$ in our simulations.

{\it Nature of disorder} \--- In the following we will consider two
different cases: (i) an annealed disorder where after a plastic slip new values of the thresholds $\sigma_{ij}^+$, $\sigma_{ij}^-$ are computed in the absence of any memory;
(ii) a quenched disorder, as a result of which the stress landscape of any given cell remains fixed so that the very same elastic branches are revisited in the course of a back and forth motion. 

The landscape with quenched disorder is implemented through the use of a counter-based random number generator (CRNG)~\cite{Shaw-SC11} so that the value of a threshold at the local elastic branch $\ell$ only depends on the index $\ell$ of that branch and on a previously defined key $\kappa$. In this way,  the access to, say, $\sigma^+_\ell=f_\kappa(\ell)$ 
requires just a simple call to the generator without the need of storing a full sequence of random numbers. 

In the following we will use an annealed disorder throughout the glass preparation step and a quenched disorder throughout the quasi-static shear driving steps. More specifically, we first ``fabricate'' our glasses using a two-step process, which mimics a thermalization step at high $T$ and a subsequent aging step at vanishing temperature. We control the degree of aging of our glasses in this manner. Further details are given in Section \ref{sec:glass-preparation}. At the end of this preparation protocol the different fields (thresholds in the forward/backward directions $\sigma^\pm_{ij}$ and  internal stress $\sigma_{ij}$) are stored; the plastic strain field is reinitialized at zero and this initial configuration is inserted as the slice of index $\ell=0$ of a stack of quenched disorder thresholds at each cell $(i,j)$. This quenched configuration is then used to perform  mechanical loading. 

\begin{figure*}
\begin{center}
\includegraphics[width=0.485\textwidth]{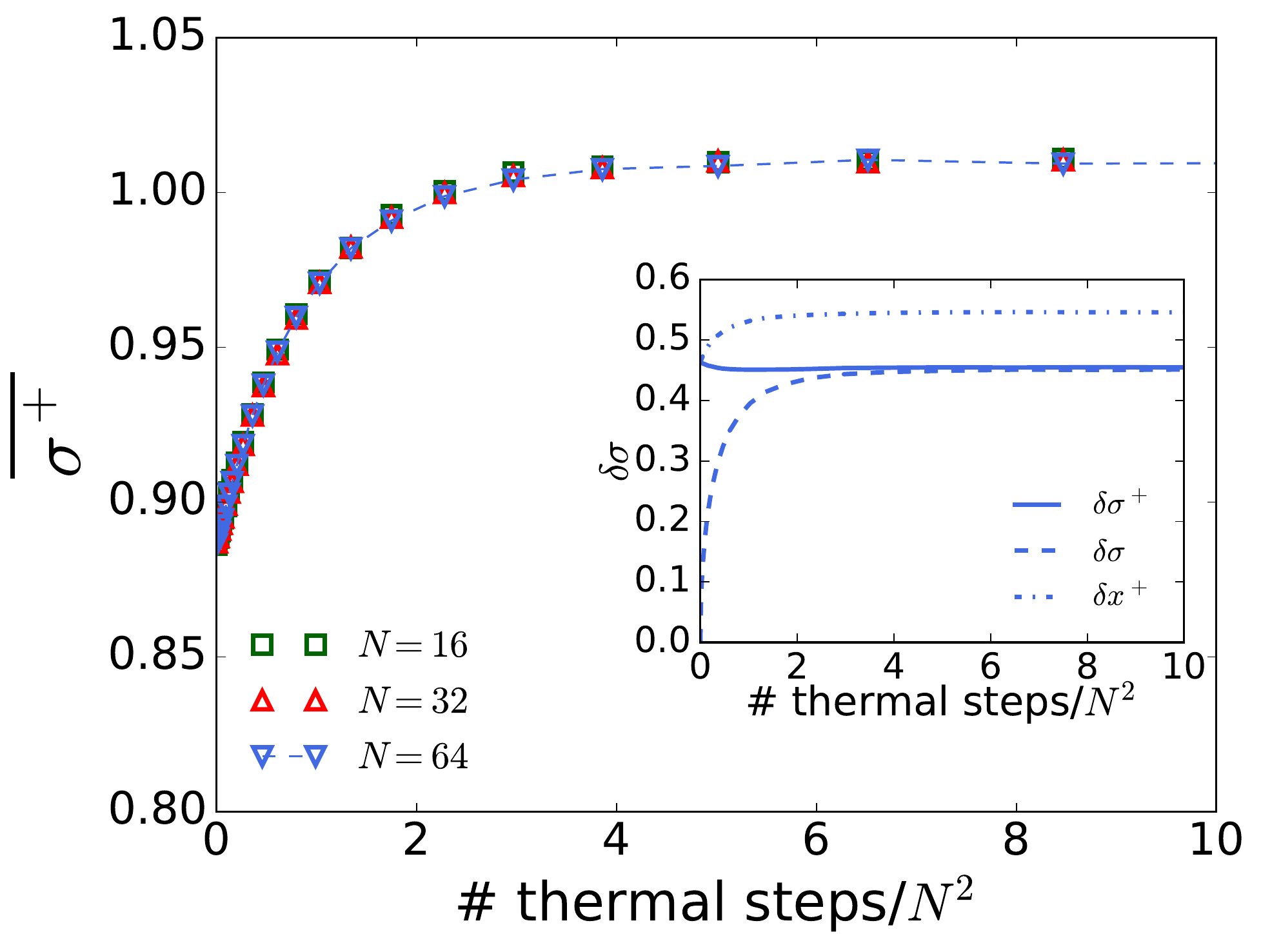}
\includegraphics[width=0.47\textwidth]{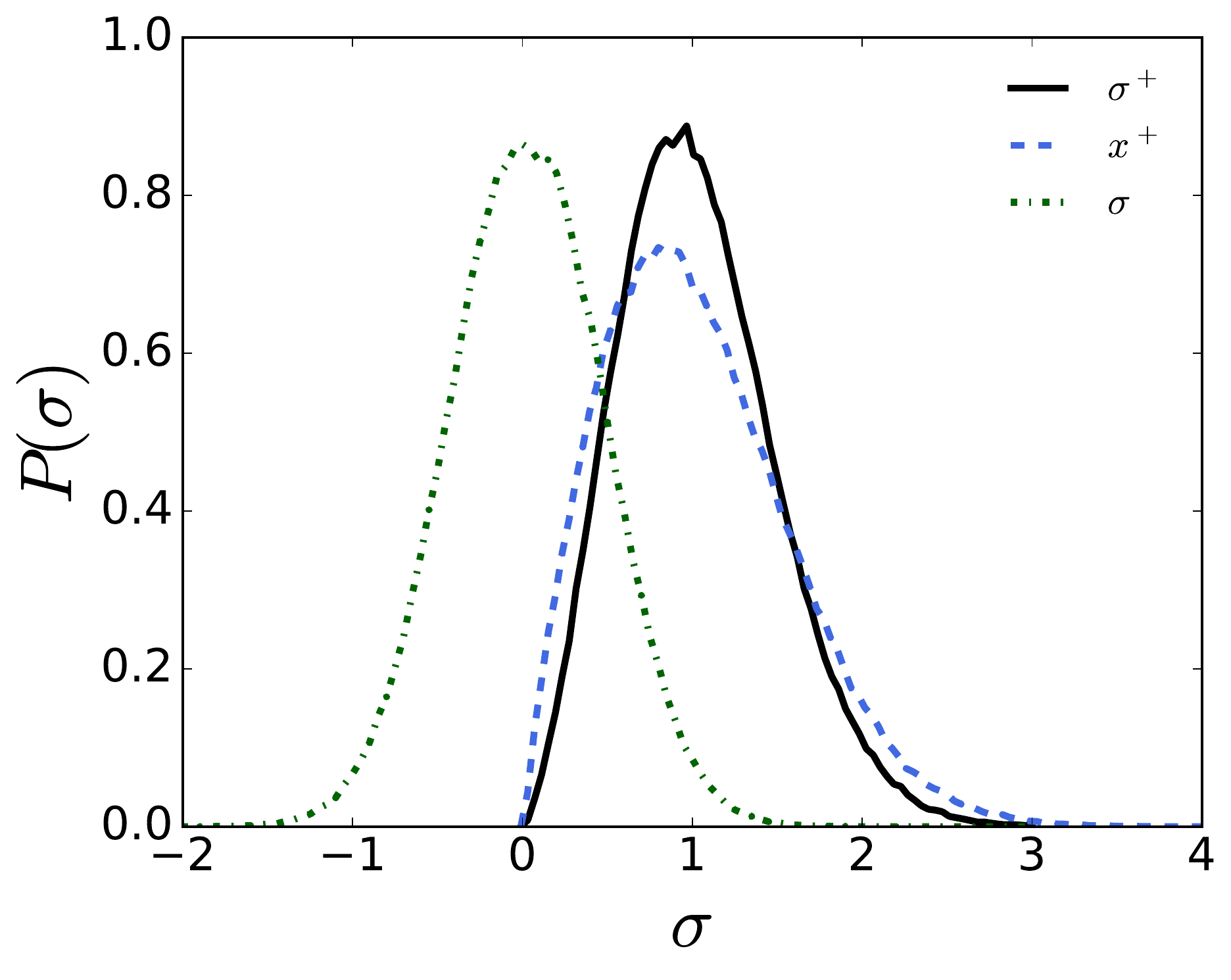}
\end{center}
\caption{\label{fig:glass-preparation-highT} Glass preparation - mimicking instant quench from high T: (a) evolution of the mean stress-threshold $\overline{\sigma^+}$ with the number of (random) thermalization steps per site and for system sizes $N = 16, 32$, and $64$. The inset shows the same evolution for the standard deviations of the stress thresholds $\delta \sigma^+$, internal stresses $\delta \sigma$, and local plastic strength $\delta x^+$. (b) stationary distributions of the fields $\sigma$, $\sigma^+$ and $x^+$ for $N = 64$. }
\end{figure*}

{\it Driving} \--- Two kinds of mechanical loading are considered in this study: monotonous shear loading and cyclic loading. In both cases, the driving is strain controlled and changed quasi-statically. The elementary steps consist in (i) identifying the first site\footnote{We will henceforth  use the terms cell and site interchangeably} $(i^*,j^*)$ which becomes unstable in the shear loading direction, i.e. the {\em extremal site}; (ii) incrementing the external strain $\varepsilon$ up to the point where the extremal site $(i^*,j^*)$  becomes unstable; 
(iii) incrementing the plastic strain of $(i^*,j^*)$ by $\Delta \varepsilon$ to trigger the transition $\ell \to \ell \pm 1$ to the next elastic branch by the instability ({\em plastification}); (iv) updating the internal stresses of all sites; (v) identifying any site that has in turn become unstable due to the internal stress update, plastifying these sites as well, updating the internal stress {\em etc.} until the end of the {\em avalanche}, i.e. until all sites have become stable again; (vi) repeat steps (i)--(v) as needed.

{\it Avalanches} \--- The precise treatment of step (v), i.e. the avalanche, deserves more detail. Once a list of unstable sites  has been identified, the question remains about the order in which these sites will be updated. Indeed, since the elastic interaction can induce both positive and negative stress kicks, an unstable site can be healed and get stable again after another one has been plastified and the resulting internal stresses at the other sites have been updated, steps (iii) and (iv). Hence the order of the updates matters. The effect of the ordering of updates on the dynamical properties has been recently discussed by Ferrero and Jagla~\cite{Jagla-SM19}. Some of us opted for a synchronous update~\cite{Tyukodi-PRL18}: all unstable sites are plastified simultaneously in parallel; the internal stress is updated afterwards; after this first sweep, a new configuration is reached, a stability test is performed, if all sites are stable, the avalanche is over, otherwise a new list of unstable sites is identified and the process is iterated until a stable configuration is reached. Here we make a different choice and perform a sequential update: the most unstable site, i.e. the extremal site,  is updated first (plastic slip followed by an update of the associated elastic stress field) and we repeat this procedure until  all sites become stable again. This choice of updating protocol happens to be very close to the extremal driving proposed in Ref.~\cite{BVR-PRL02}

\section{Glass preparation: mimicking instant quench and aging\label{sec:glass-preparation}}
\label{sec:GlassAging}

As explained before, the present model is stress based and relies on threshold dynamics: plasticity sets in at cell $(i,j)$, if and only if the local stress overcomes one of the two thresholds in the positive or negative shear directions: $ \sigma_{ij} > \sigma_{ij}^+ $ or $ \sigma_{ij} < - \sigma_{ij}^- $.
Despite the absence of an explicit energy landscape, which would allow us to equilibrate the system at finite temperature and to subsequently perform a quench to zero temperature~\cite{bulatov_stochastic_1994}, it is possible to implement two limit-cases of glass preparation: instant quench from a high temperature liquid and aging at vanishing temperature, respectively.

\begin{figure*}[t]
\includegraphics[width=0.48\textwidth]{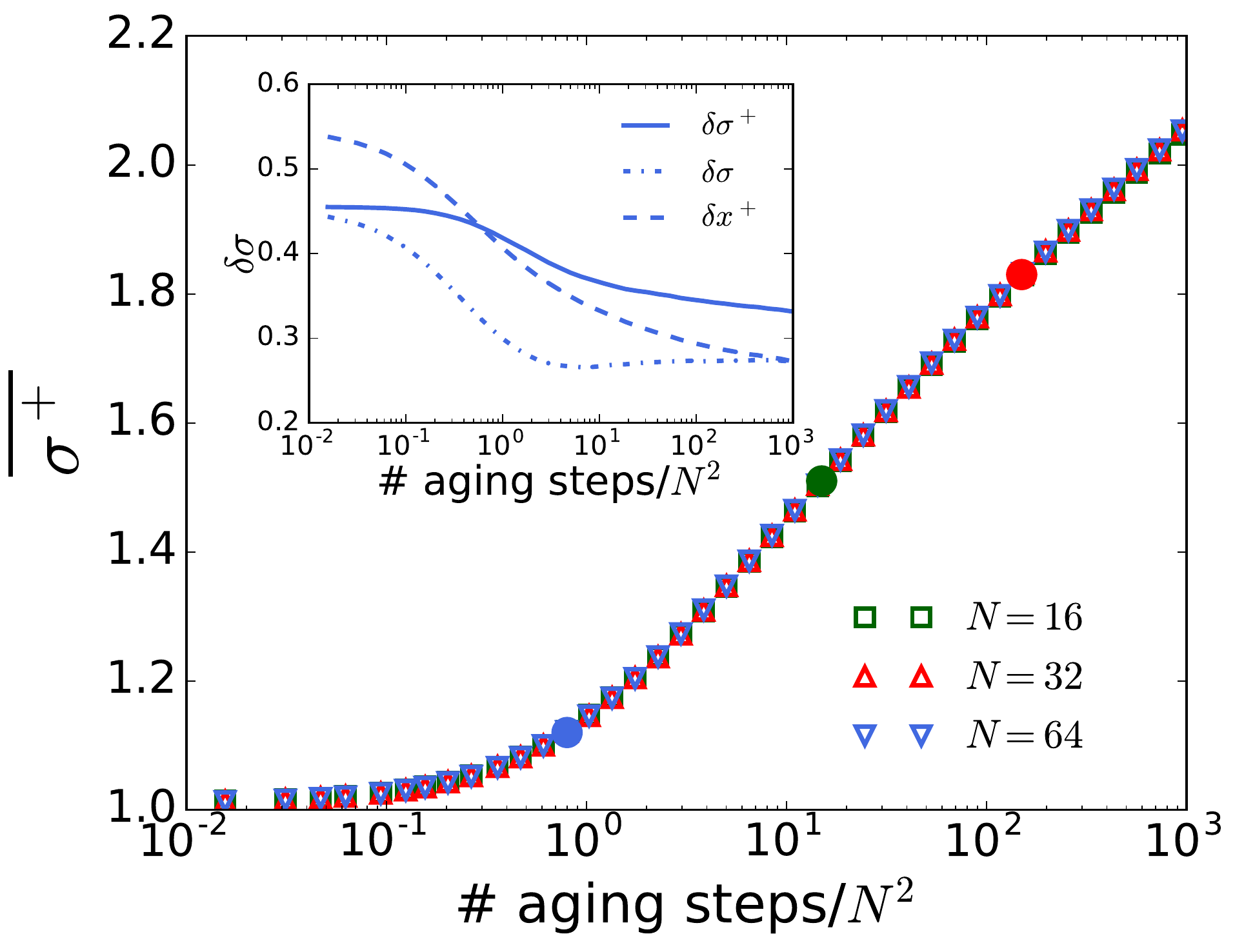}
\includegraphics[width=0.48\textwidth]{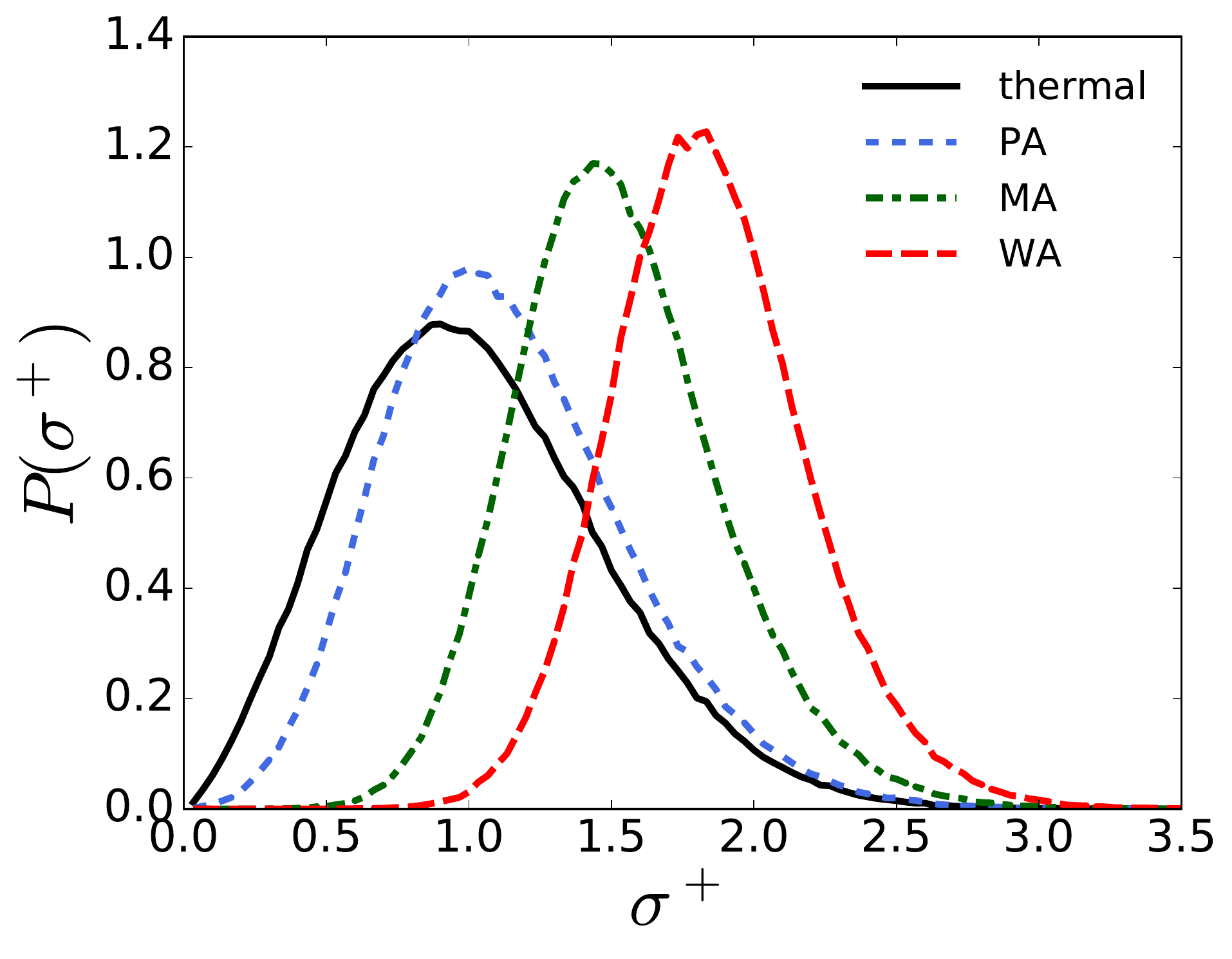}
\caption{\label{fig:glass-preparation-aging} Glass preparation - low temperature aging: (a) evolution of the mean stress-threshold $\overline{\sigma^+}$ with the number of aging steps per site and for system sizes $N = 16, 32$, and $64$.  The inset of the figure shows the evolution of the
standard deviation of local stress $\delta\sigma$, thresholds $\delta\sigma^{\pm}$ and  plastic strength $\delta x$ for $N =64$. (b) Distributions of the stress-thresholds for an $N=64$ sample, that has not been aged at all (thermal), or aged with an average number of $0.8$  (PA), $15$ (MA), and $150$ (WA) aging steps per site, corresponding to a poorly-, moderately- and well-aged glasses, respectively. The highlighted and color-coded circles in the main plot of (a) indicate the aging stages at which these samples were prepared to be subjected to cyclic shear. }
\end{figure*}

\subsection{Instant quench of a high temperature liquid}
At high temperature the local energy barriers associated  with the stress thresholds are very low with respect to the available thermal energy so that in the $T\to\infty$ limit, all plastic rearrangements are equally probable. 
We then define a {\it thermal} step by selecting a site uniformly in space at random and choosing one of the two directions with probability $1/2$.
The chosen site thus experiences a plastic slip and jumps onto a new elastic branch, which is characterized by two new plastic thresholds. Next, the stress field is updated to account for the stress redistribution. The stress redistribution can make some other sites mechanically unstable and thereby induce an avalanche. Updates are then performed until the avalanche stops and the system is stable again. The system is subjected in this manner to a sequence of thermal steps until it reaches a stationary state. In Fig.~\ref{fig:glass-preparation-highT}(a) we show for different system sizes $N$ how the mean stress-threshold $\overline{\sigma^+}$ of our samples evolves with the number of thermal steps. We see that when plotted against the average number of thermal steps per site, the curves for the different sizes collapse and $\overline{\sigma^+}$ reaches a stationary value rather quickly, after about 4-5 thermal events per site. The inset of the figure shows the corresponding evolution of the standard deviations $\delta \sigma^+, \delta \sigma$, and $\delta x^+$, of  the stress-threshold, the internal stress, and the plastic strength, respectively. When plotted against the average number of thermal steps per site, we find again little size dependence. 
In Fig.~\ref{fig:glass-preparation-highT}(b) we show the stationary distributions of the stress-thresholds, internal stress and local plastic strengths for our $N=64$ sample.

\subsection{Aging at vanishing temperature\label{subsec:aging}}

We now turn to the other limit, namely aging at very low temperature,  $T\to 0$. In the framework of activated behavior, the activity at low temperature is restricted to overcoming the lowest barriers. Moreover, in the limit of vanishing temperature, the lowest barrier becomes dominant. We define an extremal aging step as follows: recall that for each site $(i,j)$ its plastic strength in the positive and negative directions are given as $x^+_{ij} = \sigma_{ij}^+ - \sigma_{ij}$ and $x^-_{ij} = \sigma_{ij} + \sigma^-_{ij}$, respectively. We identify the site and direction with lowest plastic strength and let it experience a local slip so that stresses are redistributed, and new 
stress thresholds are assigned to the yielded site.  As in the case of the ``thermal'' procedure with randomly selected sites, a stability check is performed after each slip. If one or more sites get unstable, they are updated in turn and with the most unstable sites updated first, as explained before. The procedure is iterated until the avalanche triggered by the initial extremal step terminates. Then, the next site and direction of lowest plastic strength is identified and allowed to slip. 

The present ``aging'' procedure is thus similar to the ``thermal'' procedure, differing only in  the choice of the initial site to be slipped: in the case of ``aging'' an extremal site is selected for slip, i.e. the cell and direction with least plastic strength, while in the thermal case the selection of site and direction is random.
This difference drastically alters the dynamics, since it induces a systematic statistical bias. When a site yields, it acquires a new pair of thresholds. The latter are drawn from a prescribed distribution. But in the framework of the aging procedure this takes place at an extremal site, which is characterized by a very low plastic strength (either in the positive or in the negative direction). We thus get a typical exhaustion phenomenon: low thresholds get replaced by ``normal'' ones. This systematic bias induces a drift in the threshold distributions and thus a systematic plastic hardening~\cite{BVR-PRL02,Talamali-CRM12}. 

Starting from an initial state corresponding to the inherent state  obtained from a ``high temperature liquid'', as described in the previous section, we thus ``age'' the system by slipping a number of least stable sites. As shown in Fig.~\ref{fig:glass-preparation-aging}(a), we observe a logarithmic growth of the mean thresholds $\overline{\sigma^+}$ with the number of aging steps. Again, the dependence of this evolution on system size becomes negligible when we consider the average number of aging steps per site, instead of the total number of steps. We find that after about $10^3$ aging steps per site, the mean threshold doubles in value.

The inset of the figure shows the evolution of the standard deviation of the stress-threshold, internal stress and plastic strength. The standard deviation of thresholds shows a slow decrease (about $20\%$ over $10^3$ aging steps per site). Together with the doubling of the mean thresholds over the same range of $10^3$ steps, this corresponds to a significant narrowing of the threshold distributions upon aging.

Interestingly, after a fast decrease in the early stage of the aging protocol (less than one aging step per site) the standard deviation of internal stress remains almost constant upon aging. In recent studies
on the dependence of plastic behavior of amorphous solids on glass preparation~\cite{ozawa2018random,Rossi-arxiv22}, the width of the stress fluctuation distribution has been used as a proxy for the level of stability of the amorphous solids while keeping constant (actually uniform) the value of the plastic threshold. We get here a different situation: an increase of the mean threshold and stability of the stress fluctuations upon aging. A  way to reconcile these contrasting observations is to consider the fluctuations of the local plastic strength $x^\pm = \sigma^\pm \mp \sigma$ and to note that in the case of uniform thresholds the standard deviation of plastic strength equals that of internal stress $\delta x^\pm = \delta \sigma$. Upon aging, we indeed observe a continuous decrease of $\delta x^\pm $ which gets halved after about $10^3$ aging steps per site.

In Fig.~\ref{fig:glass-preparation-aging}(b) we display distributions of the stress thresholds $\sigma^+$ for our $N=64$ samples, which were either not aged at all (thermal), or aged at $0.8, 15$, and $150$ aging steps per site, for $N=64$. These aging levels have been indicated by the appropriately colored circles on the graph showing the evolution of mean stress-thresholds with aging in panel (a). Henceforth we will refer to these levels of aging as poorly-aged (PA), moderately-aged (MA), and well-aged (WA).

The effect of our aging procedure is dramatic: it opens a growing gap in the distribution of stress-thresholds $\sigma^+$. In spirit, we recover here a phenomenology which is close to that of ultrastable glasses obtained via swap Monte-Carlo methods~\cite{swap}. The opening of a gap will induce a  perfect elastic behavior over a finite range of strains which contrasts with the quasi-elastic behavior (short elastic branches punctuated by plastic events) typically observed in less equilibrated glasses.   

\section{Monotonous loading: dependence on thermal history\label{sec:monotonous-loading}}

Depending on glass preparation, stress-strain curves show either a monotonous behavior up to a plateau or exhibit a {\em stress peak} followed by a {\em softening branch} that slowly approaches the {\em stress plateau} at a {\em steady-state stress} $\Sigma_{\rm ss}$. 
The existence of a stress peak is usually associated with shear-banding
behavior. 

\begin{figure}
  \centering
\includegraphics[width=0.9\columnwidth]{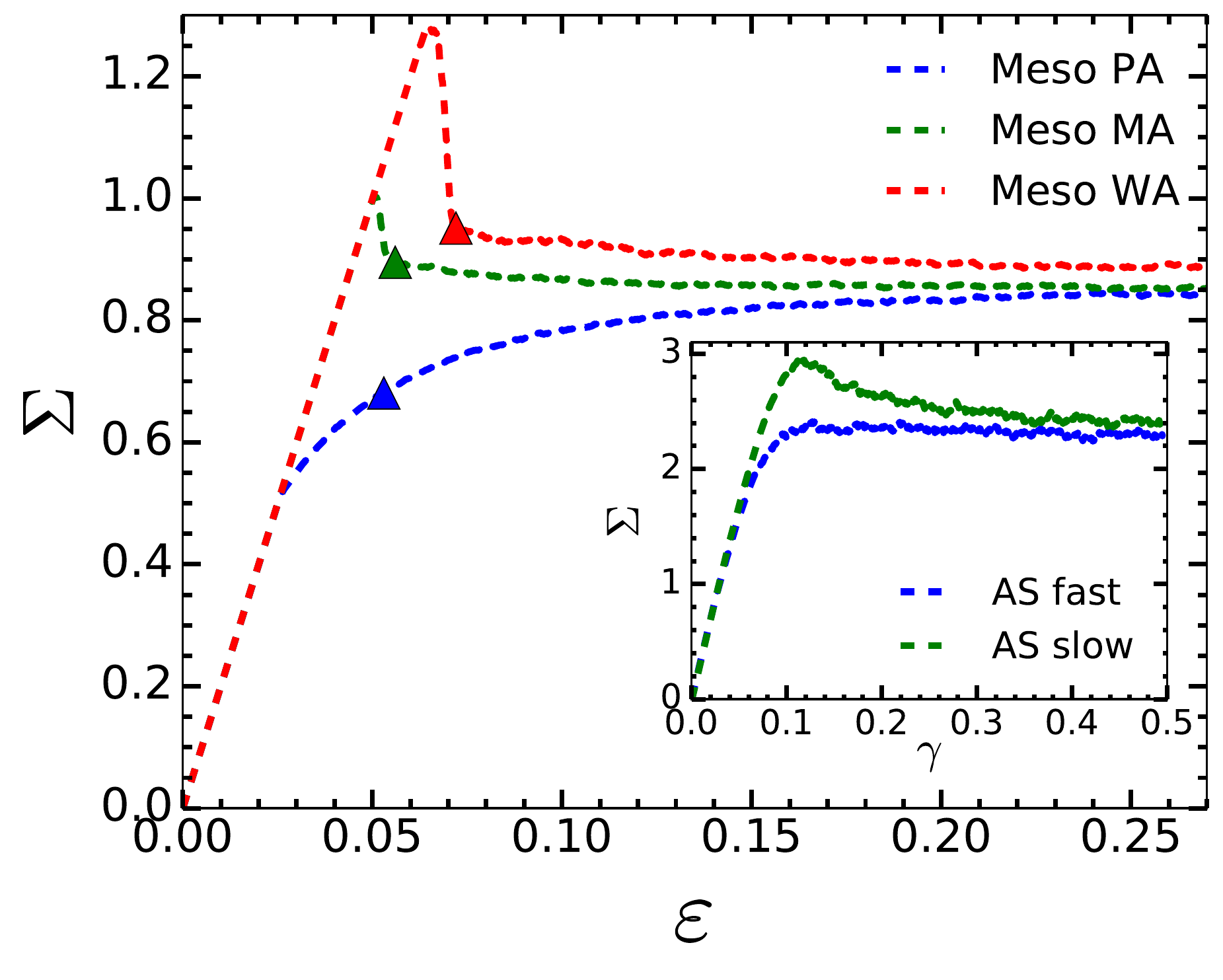}
\caption{\label{fig:monotonous-loading}Stress-strain curves upon monotonous loading. The main figures shows the stress-strain curves obtained for a mesoscopic glasses of size $N = 32$, aged at an average number of $0.8$ (poorly-aged PA), $15$ (moderately-aged MA) and $150$ (well-aged WA) aging steps per site. The moderately- and well-aged glasses show a stress peak followed by a softening branch which crosses over into a stress  plateau.  
Symbols superposed mark the strain amplitudes where the probability to find cyclic response under symmetric oscillatory shear 
is still larger than $2\%$ (refer to Section \ref{sec:cylic-driving} for details). 
The inset shows the corresponding curves obtained from simulations of 2d atomistic glasses that were quenched from a high temperature liquid state at a fast and slow rate (refer to Appendix \ref{sec:atomistic-simulations} for simulation details). 
}
\end{figure}

In section~\ref{sec:glass-preparation}, we proposed a glass preparation protocol for our mesoscopic model which mimics aging at vanishing
temperature. While tuning an aging duration is very different from
tuning a quench rate from the liquid state, both methods allow us
to transit continuously from a soft/poorly equilibrated glass to a
hard/well equilibrated glass. Our protocol actually allows us 
to obtain in this way very different glassy states. In Fig.~\ref{fig:monotonous-loading},
we show stress-strain curves corresponding to a poorly-aged, a medium-aged, and well-aged  glass, aged at an average of $0.8, 15$, $150$ number of steps per site. The system size is $N = 32$ and the curves were obtained by averaging over 500 realizations.
While the poorly-aged glass does not exhibit a stress peak, such a peak emerges and becomes more pronounced as the samples are aged more. Thus by tuning the duration of aging we are able to transit from a poorly-aged to a well-aged glass. More details on the size dependence of these stress-strain curves are shown in Fig.~\ref{fig:meso_uni_shear} of Appendix~\ref{sec:meso_uniform_shear}.

{\it Comparison with atomistic simulations} \--- In the inset of Fig.~\ref{fig:monotonous-loading}, we show for reference two
stress-strain curves obtained by atomistic simulations under athermal
quasi-static shear for a slow and fast quench, respectively. The details of the simulations are provided in Appendix \ref{sec:atomistic-simulations}. The
slow quench curve shows a distinct stress peak while apart from
fluctuations, the fast quench curve is almost monotonous. Due to
computational time limitations, it is difficult to obtain strongly
contrasting quenches and consequently stress-strain curves when using
molecular dynamics for the glass preparation. The recently developed
swap Monte-Carlo methods give access to a wider range of glass
preparation although they are more restrictive with respect to the
nature of the model glasses~\cite{swap}. 

Let us emphasize that it has not been attempted here to
adjust the parameters of the elasto-plastic model to quantitatively
reproduce the stress-strain curve obtained by atomistic
simulations. Rather, our goal is to compare generic features,
such as the brittle to ductile transition under monotonous loading, and how the behavior upon cyclic loading depends on the soft/hard
nature of a glass. Recent analyses of coarse-graining atomistic
simulation to be used to feed mesoscopic elasto-plastic models with realistic
parameters can be found in Ref.~\cite{Patinet-CRP21}.

\section{Cyclic driving: limit cycles\label{sec:cylic-driving}}

In this section we consider the irreversibility transition, and in particular the response to cyclic shear of our poorly-aged (PA) and moderately-aged (MA) mesoscopic glasses whose preparation was described in section \ref{sec:glass-preparation}. The well-aged (WA) mesoscopic glasses yield a response to cyclic shear that is qualitatively similar to that of the (MA) glasses and will therefore not be considered in this section.

\begin{figure}
\includegraphics[width=\columnwidth]{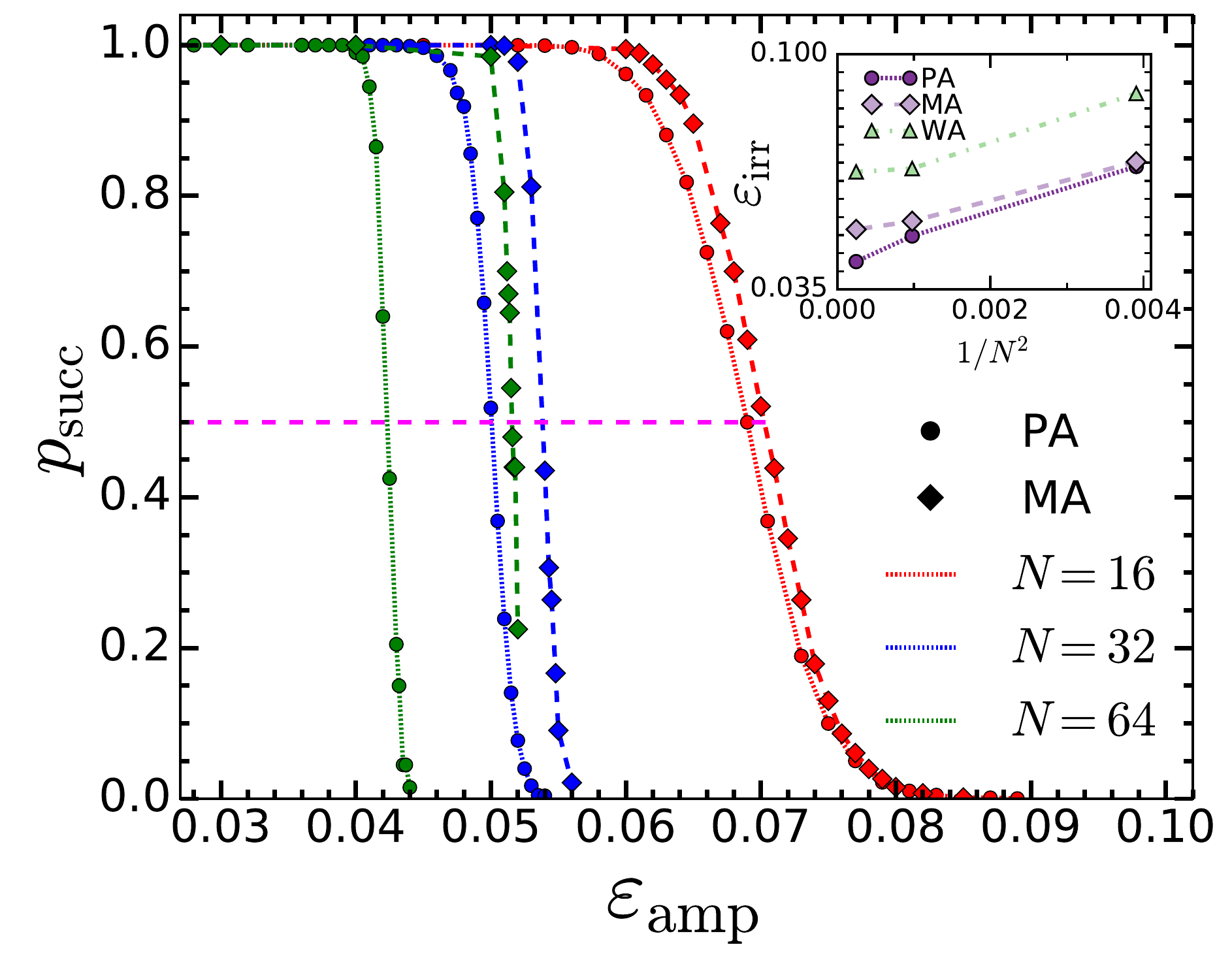}
\caption{Success rate $p_{\rm succ}$ of the convergence to a limit cycle under cyclic shearing at amplitude $\epsilon_{\rm amp}$. Shown are results for ensembles of poorly-aged (circles)  and moderately-aged (diamond) glasses with system sizes $N = 16$ (red), $32$ (blue), and $64$ (green). Intersections with the dashed horizontal line indicate strain amplitudes where the probability of finding a limit-cycle is $1/2$. 
Inset: The plot of strain amplitudes $\epsilon_{\rm irr}$ at which $p_{\rm succ} = 1/2$ against $1/N^2$ for the poorly-, moderately-, and well-aged glasses, PA, MA and WA.
}
\label{fig:success_rate}
\end{figure}

\subsection{Irreversibility transition\label{subsec:irreversibility-transition}}

\begin{figure}
\includegraphics[width=0.9\columnwidth]{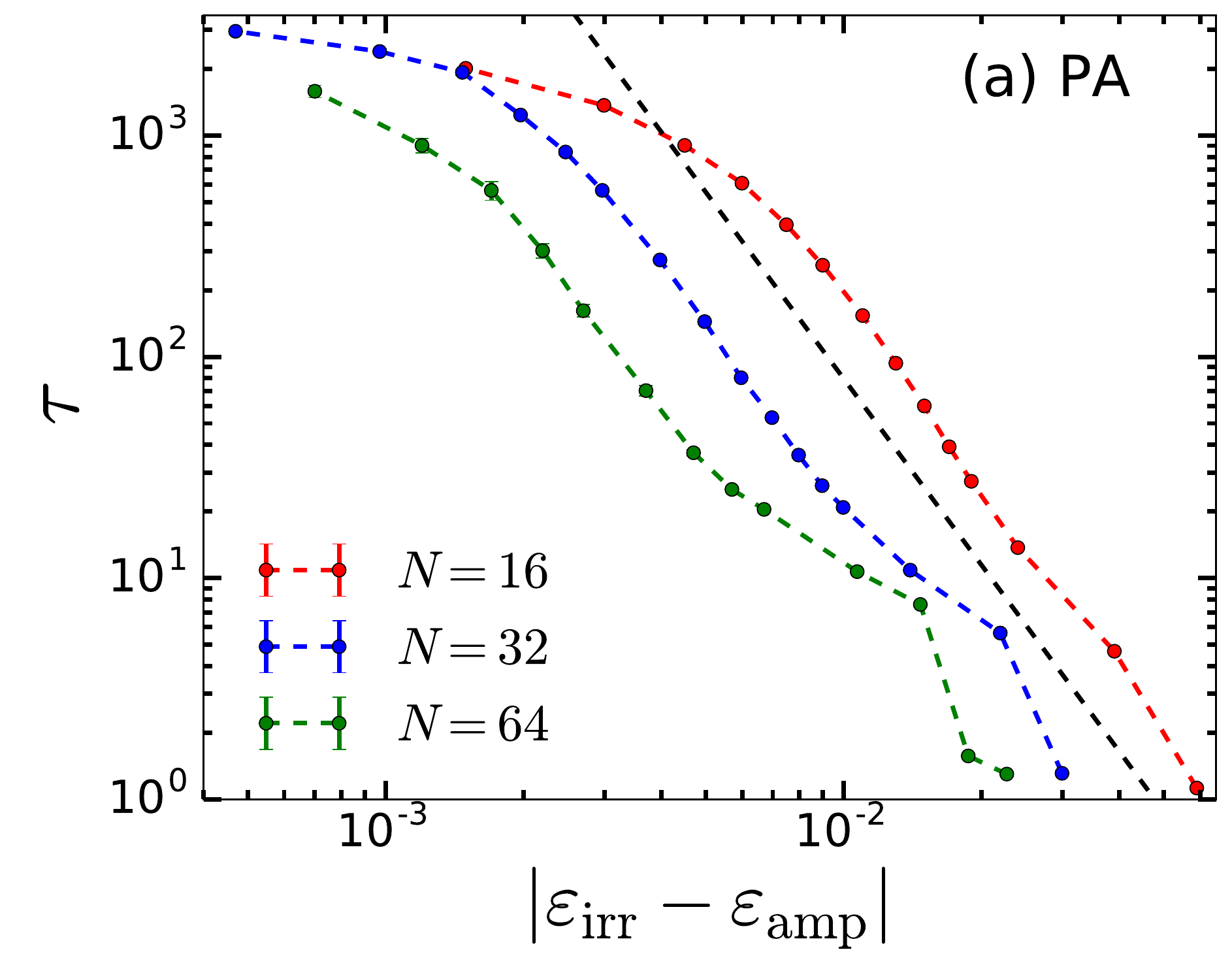}
\includegraphics[width=0.9\columnwidth]{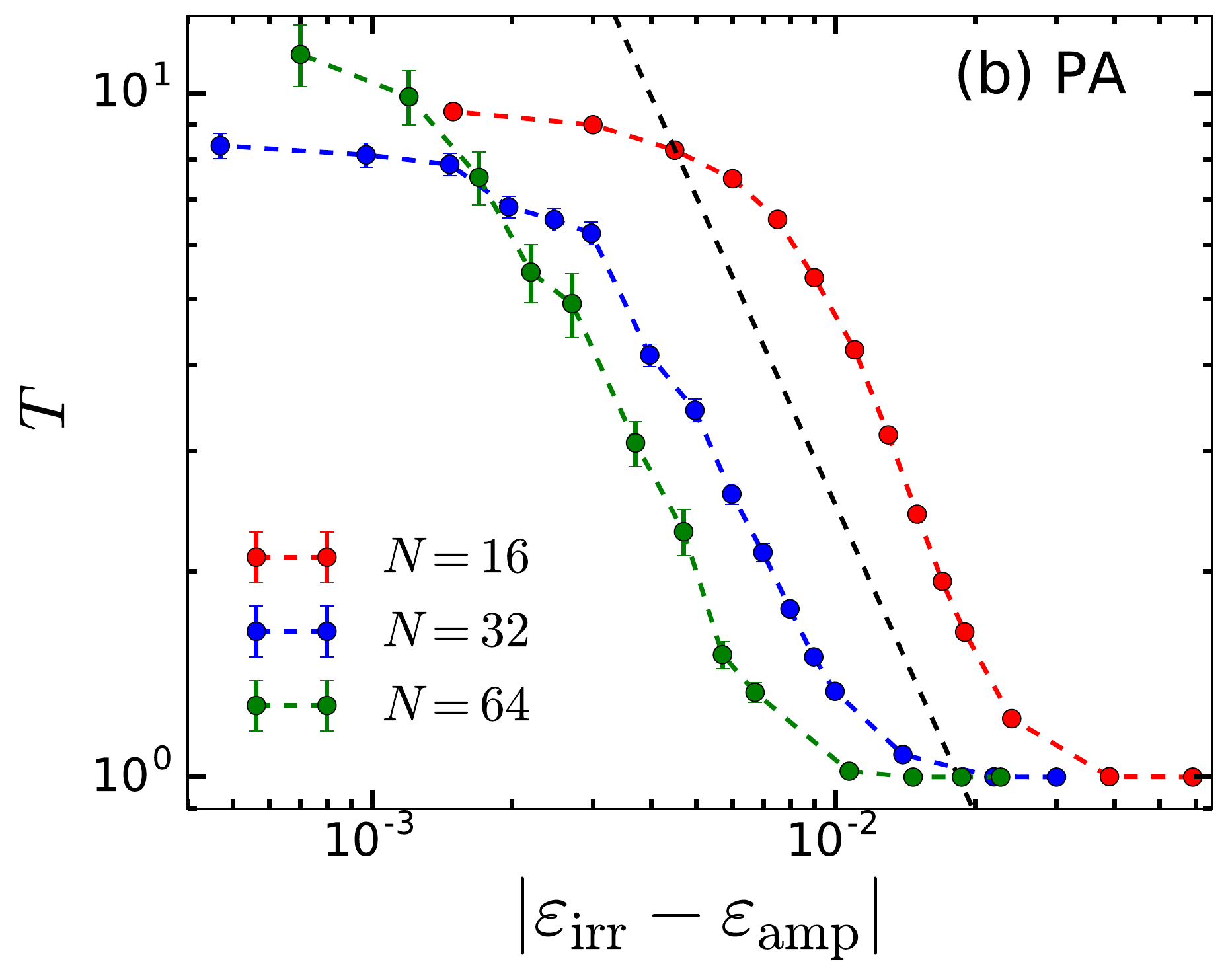}
\caption{\label{fig:transient} Convergence to limit cycles of poorly-aged (PA) glasses: (a) Duration $\tau$ of transients {\em vs.} relative cycle amplitude $\vert \epsilon_{\rm irr} - \epsilon_{\rm amp} \vert$, where $\epsilon_{\rm irr}$ is the system size dependent strain amplitude where the success-rate $p_{\rm succ}$ of cyclic response is $1/2$, {\em cf.} Fig.~\ref{fig:success_rate}. The dashed line is a power-law with exponent $2.7$ and serves as a guide to the eye. 
(b) The period $T$ of the cyclic response in units of the number of driving cycles for the poorly-aged samples at different system sizes. The dashed line is a power-law with exponent $1.5$ and serves as a guide to the eye. 
}
\end{figure}

\begin{figure}
\includegraphics[width=0.9\columnwidth]{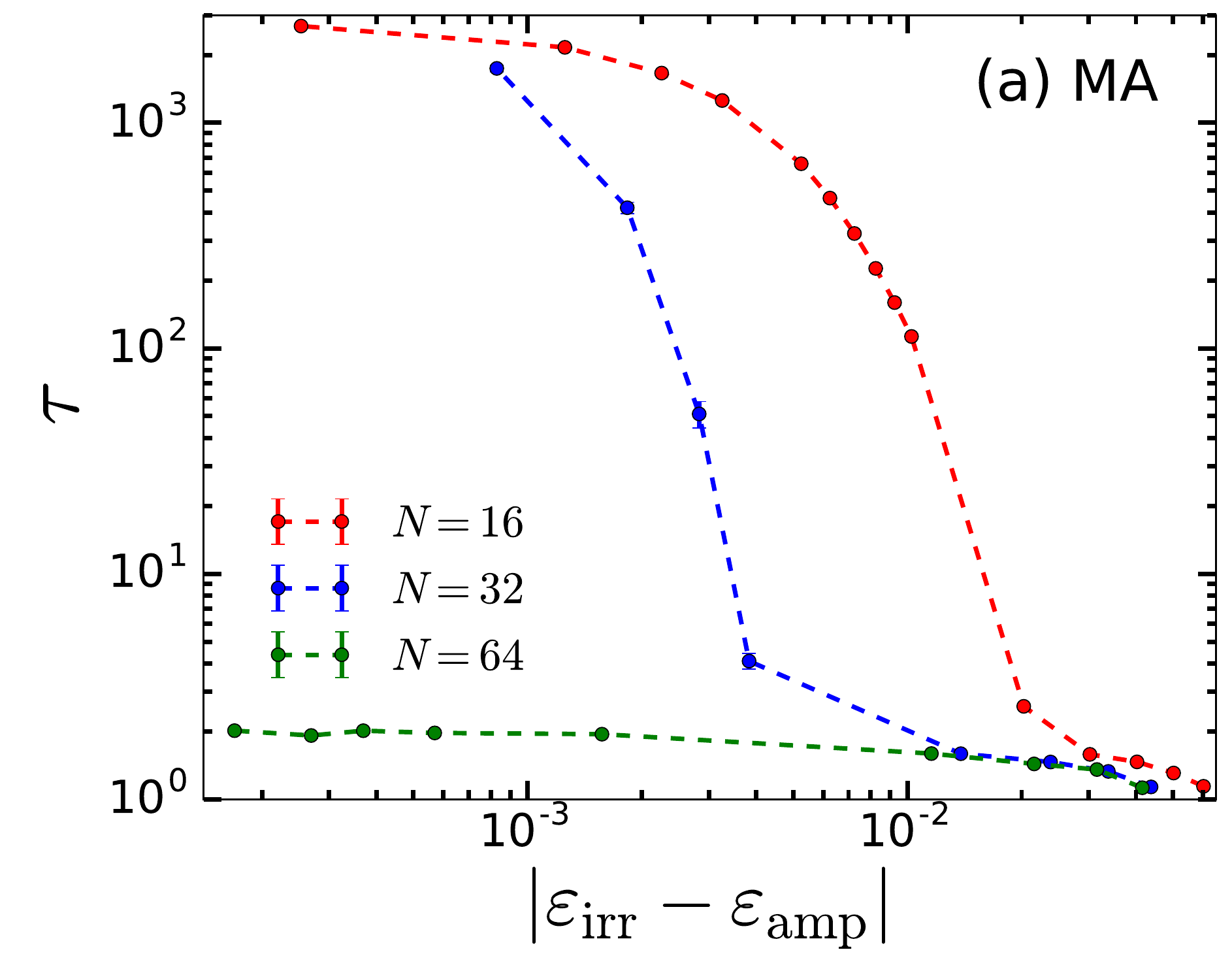}
\includegraphics[width=0.9\columnwidth]{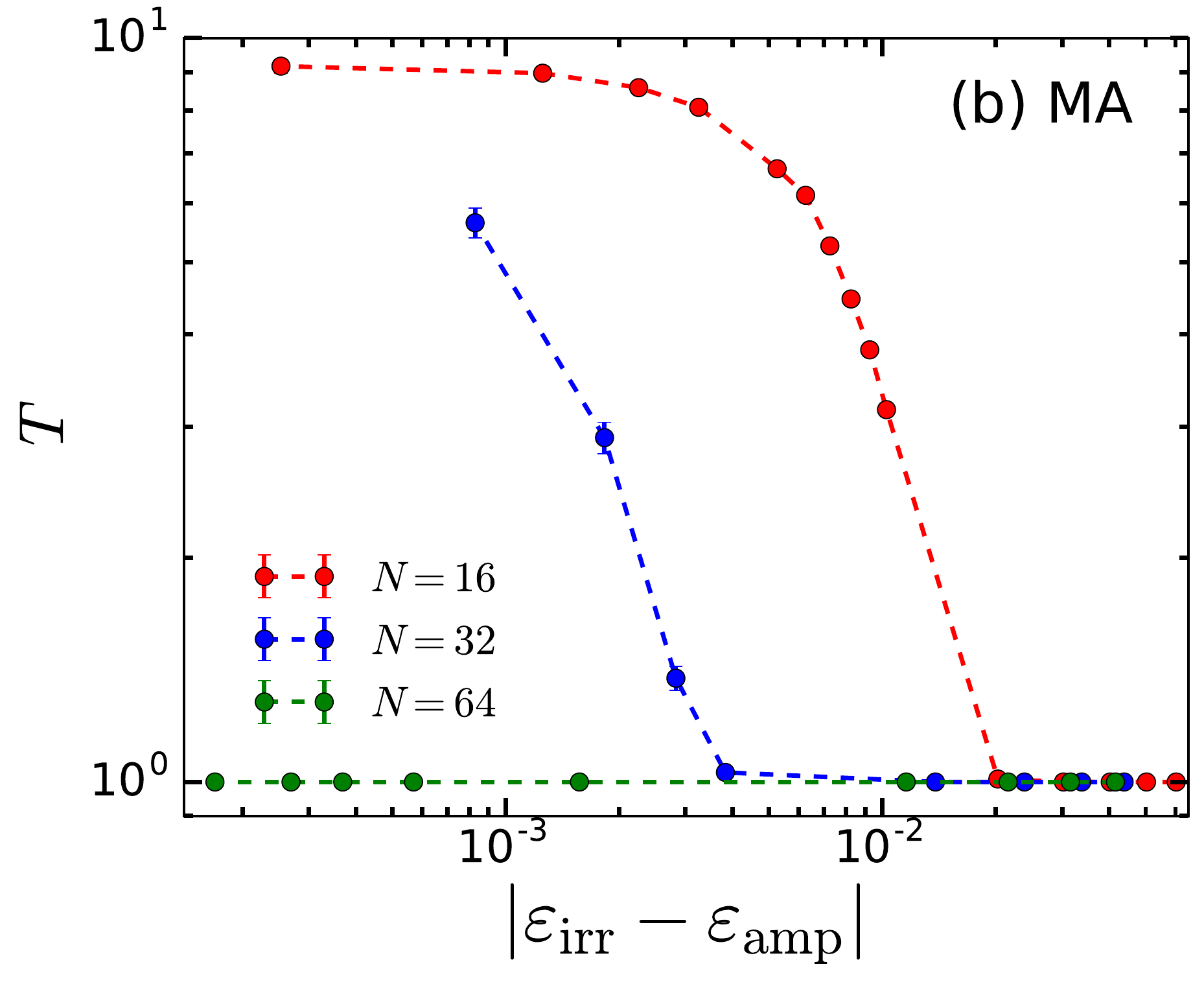}
\caption{\label{fig:transientMA} Convergence to limit cycles of moderately-aged (MA) glasses: (a) Duration $\tau$ of transients {\em vs.} relative cycle amplitude $\vert \epsilon_{\rm irr} - \epsilon_{\rm amp} \vert$, where $\epsilon_{\rm irr}$ is the system size dependent strain amplitude where the success-rate $p_{\rm succ}$ of cyclic response is $1/2$, {\em cf.} Fig.~\ref{fig:success_rate} (inset).
(b) The period $T$ of the cyclic response in units of the number of driving cycles for the moderately-aged samples at different system sizes. 
}
\end{figure}

When subjected to cyclic shear loading, amorphous solids tend to either evolve into periodic response or reach a diffusive regime, depending on the value of the amplitude $\epsilon_{\rm amp}$
of the loading cycles. This transition presents typical features of a critical transition. In particular, power-law divergence of the number $\tau$ of loading cycles to reach the periodic response below the transition, as well as the power law dependence of the diffusivity above the transition have been observed both for atomistic and mesoscopic models~\cite{fiocco2013oscillatory,regev2013onset,PRIEZJEV2013,regev2015reversibility,leishangthem2017yielding,parmar2019strain,yeh2020glass,bhaumik2021role}. The features of the irreversibility transition 
depend on glass preparation~\cite{bhaumik2021role,yeh2020glass,bhaumik2022avalanches,bhaumik2022yielding}.
Here we show results for the size dependence of the irreversibility transition in our PA and MA mesoscopic glasses. Specifically, we consider systems of size $N = 16 (7500), \ 32 (1500), \ 64 (200)$, where the numbers in parenthesis indicate the number of realizations used to obtain our results.

We first focus on the poorly-aged  (PA) systems.  Figure~\ref{fig:success_rate} shows the mean {\em success-rate $p_{\rm succ}$, i.e.} the fraction of PA systems (circles) within our ensemble of realizations that reach a limit cycle when subject to a given number $\tau_{\rm max}$ of {\em symmetric} loading cycles at amplitude $\epsilon_{\rm amp}$: $0 \to \epsilon_{\rm amp} \to 0 \to - \epsilon_{\rm amp} \to 0$. The different colors correspond to the system sizes, as indicated in the legend of the figure. For system sizes $N = 16$ and $32$ we used a cut-off of $\tau_{\rm max} = 10^4$ driving cycles, so that if cyclic response had not been established at that point we considered the run to be unsuccessful. For the $N=64$ sample this cut-off was chosen to be $\tau_{\rm max} = 5.10^3$. A clear transition can be observed between a low amplitude regime with convergence to a limit cycle and a high amplitude regime with no limit cycle. The transition between these two regimes gets increasingly sharper with  system size. A clear size dependence is also observed in the location of the transition which tends to occur at lower strain amplitudes for larger systems. 
The size effect exhibited by our poorly-aged glasses is all the more striking as it turns out to be completely absent in the response to monotonous loading, and only weakly present in the case of our moderately- and well-aged glasses,  (Fig.~\ref{fig:meso_uni_shear}  in Appendix \ref{sec:meso_uniform_shear}).  

For each size $N$, we estimate the strain $\epsilon_{\rm irr}(N)$ at which the irreversibility transition occurs, as the loading amplitude such that $50\%$ of the realizations reach a limit cycle, i.e. $p_{\rm succ} = 1/2$, as indicated by the pink horizontal line in Fig.~\ref{fig:success_rate}. The inset of Fig.~\ref{fig:success_rate}  shows the size and ageing dependence of $\epsilon_{\rm irr}(N)$ for $N = 16, 32$, and $64$, for the PA, MA and WA glasses. 
We see that for a given degree of ageing, $\epsilon_{\rm irr}(N)$ decreases with increasing system size. Moreover, a dependence of $\epsilon_{\rm irr}$ on aging at fixed system size is clearly visible, in particular for the larger sizes $N = 32$ and $64$. At these sizes the MA glasses have slightly larger $\epsilon_{\rm irr}$ then the PA ones, while the WA glasses have overall larger values of $\epsilon_{\rm irr}$ for all system sizes considered. The behavior of $\epsilon_{\rm irr}$ with aging is consistent with atomistic simulations of cyclically sheared amorphous solids which show that the strain marking the onset of the irreversibility transition is largely independent of aging for sufficiently poorly-aged samples, but that it starts to increase as the samples are better aged\cite{yeh2020glass,bhaumik2021role}.

We turn next to the response of our moderately-aged (MA) glasses to cyclic shear. The diamond symbols in Figure~\ref{fig:success_rate} show the fraction $p_{\rm succ}$ of MA glasses in our ensembles of realizations that reach a limit cycle when subject to cyclic loading of amplitude $\epsilon_{\rm amp}$. Similarly to the poorly-aged samples, as the system size is increased, the irreversibility transition exhibits an increasingly sharper decline of the success-rate from one to zero.
However for a given system size, the rapid fall-off of the success rate in the MA glasses occurs at consistently larger strain values than for the PA glasses,  which is in agreement with the behavior of $\epsilon_{\rm irr}$ discussed above.

\subsection{Transient regime and limit cycles\label{subsec:transient-regime}}

Another feature of the irreversibility transition is the divergence of the duration of the transient regime: atomistic simulations show that the number of loading cycles needed to reach the limit cycles diverges as a power-law according to  $\tau(\epsilon_{\rm amp}) \propto |\epsilon_{\rm irr} - \epsilon_{\rm amp}|^{-\alpha}$, as shown in Refs.~\cite{regev2013onset,regev2015reversibility,kawasaki2016macroscopic,Maloney2021}. 

In Fig.~\ref{fig:transient}a, we plot $\tau(\epsilon_{\rm amp})$ against  $|\varepsilon_{\rm irr}(N) - \epsilon_{\rm amp}|$ for our poorly-aged glasses and different system sizes $N$. Here $\varepsilon_{\rm irr}(N)$ is the loading amplitude at which half of the realizations reach limit cycle, as defined previously. Once again, a significant size effect is observed: for a given $|\epsilon_{\rm irr}(N) - \epsilon_{\rm amp}|$, the larger the system size, the shorter the transient regime. An indicative power-law behavior of exponent $\alpha=2.7$, as recently reported in Ref.~\cite{Maloney2021}, is plotted as a dashed line. We see that the results obtained for $N=16, 32$, and $64$ are reasonably consistent with this trend over roughly one decade for the larger samples. Note that the value $\alpha=2.7$ is close to the estimate $\alpha\approx 2.6$  obtained using atomistic simulations by Regev et al.~\cite{regev2013onset, regev2015reversibility} and also close to the value $\alpha\approx 2.66$ obtained by Cort\'e et al.~\cite{corte2008random} for a simplified model of interacting particles under flow. 
The saturation observed for large values of $\tau(\epsilon_{\rm amp})$ naturally stems from the hard limit associated with the finite number of loading cycles $\tau_{\rm max}=10^4$ for $N = 16,32$, and $\tau_{\rm max}=5.10^3$ for $N=64$ that we used in our numerical  simulations. 

In Fig.~\ref{fig:transient}b, we also plot the period of the limit cycle $T(\epsilon_{\rm amp})$ against  $|\epsilon_{\rm irr}(N) - \epsilon_{\rm amp}|$ for our poorly-aged glasses and different system sizes $N$. As already observed in Ref~\cite{Maloney2021}, we see that the limit cycles get more and more complex, with an increasing period when the amplitude $\epsilon_{\rm amp}$ of the cyclic loading approaches the irreversibility transition $\epsilon_{\rm irr}$. For illustrative purpose we show that the fast increase of the period is consistent with a power law behavior  $T(\epsilon_{\rm amp}) \propto|\epsilon_{\rm irr}(N) - \epsilon_{\rm amp}|^{-\beta}$ with $\beta = 1.5$  plotted as a dashed line in Fig.~\ref{fig:transient}b. 

In Fig.~\ref{fig:transientMA}, we show the same observables $\tau(\varepsilon_{\rm amp})$ and $T(\varepsilon_{\rm amp})$ close to the irreversibility transition, now for the moderately-aged glasses. For small systems sizes ($N=16,32$), we again observe a diverging trend in the transient duration and the limit cycles period. It appears actually that the larger the system size, the narrower the range of amplitudes over which this diverging behaviour holds. Another behaviour gradually becomes dominant: for large system sizes, a limit cycle is reached after just a few loading cycles, and the response is mainly elastic. Moreover, as it can be seen for the $N=64$ glass in Figs.~\ref{fig:success_rate} and \ref{fig:transientMA} the transition to irreversibility is rather abrupt and discontinuous. The system either reaches a $T = 1$ cyclic response rather quickly or not.

\begin{figure*}
\includegraphics[width = \textwidth]{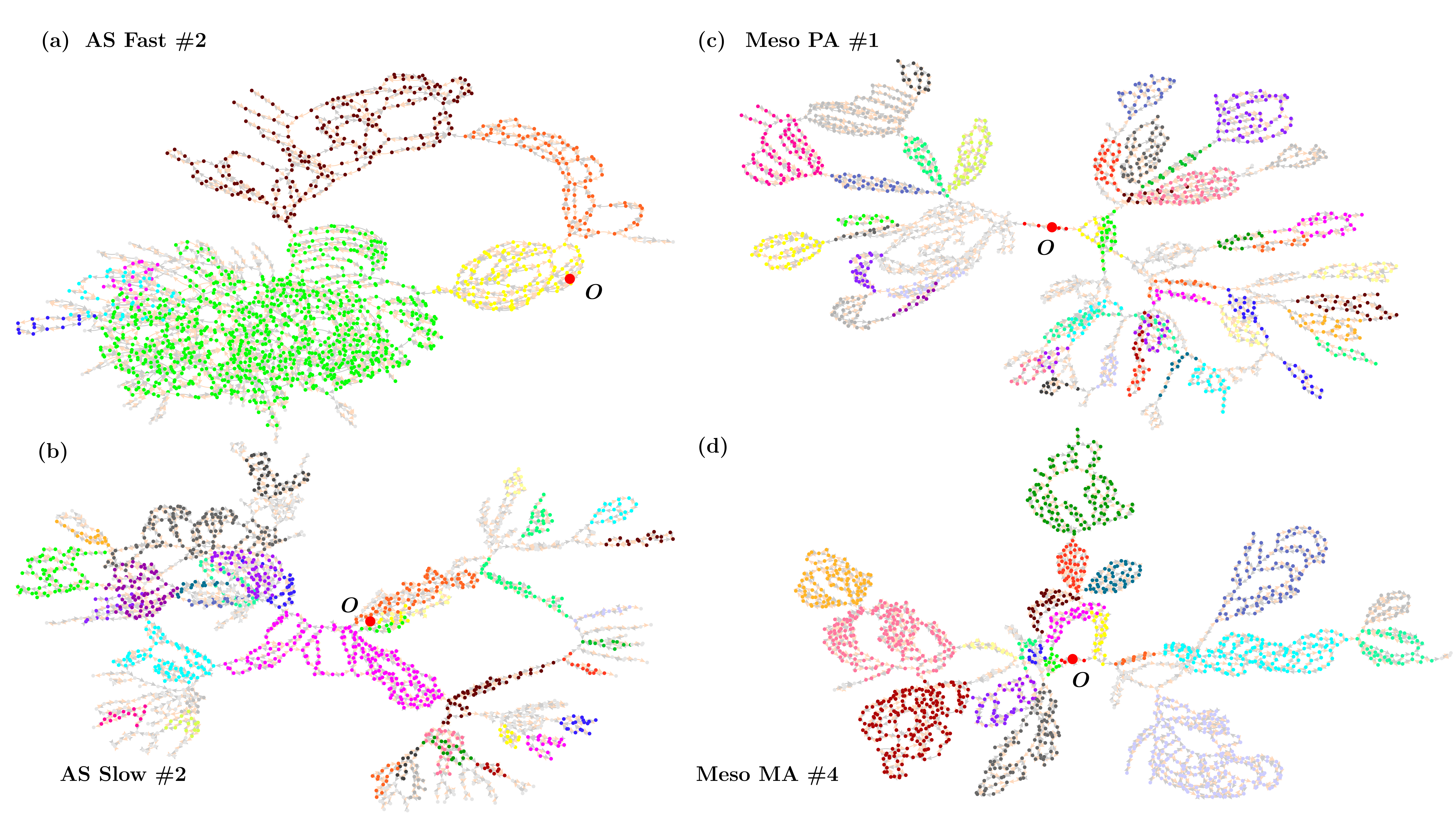}
\caption{\label{fig:graphs} Transition-graph representation of the AQS dynamics and thermal history -- atomistic (AS) {\em vs} mesoscopic (Meso) models. Excerpts of transition graphs extracted from atomistic (a,b) and $N = 32$ mesoscopic glasses (c,d) with different thermal histories: (a,c) poorly-aged/fast quenched, (b,d) moderately aged/slow quenched. The color of each vertex indicates the  strongly connected component (SCC) of the graph that it belongs to (refer to text for details) and the initial mesostate of the prepared glass has been marked with a larger red vertex labeled $\bf{O}$.
}
\label{fig:tgraphs}
\end{figure*}

\section{Characterization of the disorder 
 landscape via transition graphs\label{sec:t-graphs}} 

In order to compare the disorder landscape obtained from our mesoscopic and atomistic  simulations, we turn next to the transition graph ($t$-graph) representation of the dynamics under  AQS shear~\cite{Maloney2006}. As was shown recently~\cite{Mungan-PRL19, Regev2021}, such $t$-graphs can be extracted from atomistic simulations of sheared amorphous solids. Features of the AQS dynamics, such as yielding and return point memory, are thereby encoded in the topology of the $t$-graph~\cite{munganterzi2018structure,Mungan-PRL19,Regev2021}. Thus $t$-graphs provide useful information about the underlying disorder landscape. 
At the same time, the representation of AQS dynamics via $t$-graphs extracted from simulations provides a unified framework within which we can compare the dynamics of atomistic as well as mesoscopic models in a rather direct and comprehensive manner. This is the aim of the present section. 

\subsection{AQS transition graphs\label{AQS-t-graphs}}
To fix ideas, we consider first the sheared amorphous solid in an atomistic setting. Under AQS conditions,  a given mechanically stable particle configuration can be sheared in the positive and negative direction until a mechanical instability occurs. Denoting by $\epsilon^\pm$ 
the critical values of the external shear strain at which the instability sets in, for shear strains between $\epsilon^-$ and  $\epsilon^+$, the configuration of particles 
deforms smoothly and reversibly in response to the applied shear strain. These sets of mechanically stable particle configurations constitute an elastic branch of the system which we simply refer to as a {\it mesostate}~\cite{Mungan-PRL19}. We will use capital letters to label mesostates, and denote the critical strain values of a mesostate $A$  by $\epsilon^\pm[A]$. When 
$\epsilon = \epsilon^+[A]$ (or $\epsilon = \epsilon^-[A]$), a fast relaxation to a new mechanically stable particle configuration occurs. This particle configuration must necessarily be part of another mesostate, i.e. belong to a different elastic branch, say $B$. Thus the instability at $\epsilon = \epsilon^+[A]$ triggers a transition from mesostate $A$ to $B$. A similar transition occurs when $\epsilon = \epsilon^-[A]$. The transition between mesostates can therefore be represented in terms of a directed graph, the AQS transition graph or simply $t$-graph. The vertices of the $t$-graph are the mesostates, while from each mesostate we have two outgoing transitions which constitute the directed edges of the graph. We shall denote the  transitions  when $\epsilon = \epsilon^+[A]$ or $\epsilon = \epsilon^-[A]$ as the $\Up$-, respectively, $\Dn$-transition out of $A$, referring to the states that these lead to as $\Up A$ and $\Dn A$. 

The $t$-graph along with the critical strains $\epsilon^\pm[A]$ associated with each mesostate forms a complete representation of the AQS dynamics under arbitrary shearing protocols~\cite{munganterzi2018structure}. Given an initial mesostate $A$ and a shear protocol, the sequence of mesostates visited can be read off by following the corresponding $\Up$- and $\Dn$-edges, while checking each time whether the critical strains needed to trigger the transition have been exceeded or not.    

Note that since $\Up A$ and $\Dn A$ are mesostates reached from $A$, their stability ranges must contain the strains $\epsilon^\pm[A]$ at which these transitions were triggered, i.e. we have the AQS conditions~\cite{munganterzi2018structure}
\begin{align}
    \epsilon^-[\Dn A] &< \epsilon^-[A] < \epsilon^+[\Dn A], \nonumber \\
    \epsilon^-[\Up A] &< \epsilon^+[A] < \epsilon^+[\Up A].
\end{align}
It thus follows that $\epsilon^+[A] < \epsilon^+[\Up A] < \epsilon^+[\Up^2 A] < \ldots $ and thus the upper critical strains are monotonously increasing with repeated $\Up$-transitions. An analogous result holds for the lower strain threshold under $\Dn$-transitions. An immediate consequence of this observation is that the sub $t$-graphs, which are obtained by considering only transitions under $\Up$ (or $\Dn$), are necessarily acylic, i.e. they cannot contain any cycles. Thus any cyclic behavior must arise from an interplay of the $\Up$- and $\Dn$-transitions.

\subsection{Catalog acquisition and $t$-graphs from simulations\label{subsec:catalogs}}

The numerical algorithm of extracting $t$-graphs from simulations of sheared amorphous solids has been described in detail in the supplementary material of Ref.~\cite{Mungan-PRL19}. Here we will sketch out the main idea. We start with an initial configuration that is part of a mesostate $O$ which we call the reference state and assign it a generation number $g = 0$. Next, we execute the $\Up$- and $\Dn$-mesostate transitions out of $O$, leading to the mesostates  $\Up O$ and $\Dn O$, which we assign to generation $g = 1$. Everytime we reach a new mesostate, we compare it to the catalog of mesostates we have obtained so far to see whether it has been encountered before. If not, we add it to our catalog. By proceeding generation by generation, we acquire in this manner a catalog of mesostates: each mesostate $A$ is assigned an ID, its critical strains $\epsilon^\pm[A]$ 
and the IDs of the mesostates it transits into under a $\Up$- or $\Dn$-transitions are determined. 
The $t$-graph is then assembled from such catalogs. 
In our mesoscopic models, each mesostate corresponds to a configuration of the local elastic branches associated with each of the cells. The event based nature of their simulations facilitates the identification of mesostates and their transitions. 

We obtain catalogs from $10$ realizations each of the $N = 32$ poorly-, moderately-, and well-aged glasses, 
as described in the previous section. In addition, we produced $10$ catalogs from samples of an ultra-stable glass aged by an average of $4.10^3$ steps per-site.  
For comparison purposes, we also extracted catalogs from our atomistic simulations, using a set of $8$ soft and $30$ moderately hard reference configurations, that were obtained via fast and slow quenches from a high-temperature liquid. 
The description of these atomistic catalogs is given in  Appendix \ref{sec:catalog-properties}. 

Fig.~\ref{fig:tgraphs} shows sample $t$-graphs from each of the four sets of samples: fast quenched atomistic glass (AS Fast \#2), slow quenched atomistic glass (AS Slow \#2), poorly-aged mesoscopic glass (Meso PA \#1), and the moderately-aged mesoscopic glass (Meso MA \#4). The numbers after the $\#$ sign specify the particular realization of the glass, as listed in Tables \ref{tab:N32PAcatalogs}, \ref{tab:N32WAcatalogs}, \ref{tab:glv2catalogs}, and \ref{tab:waglv2catalogs}. The placement of the vertices of the graph is arbitrary. The mesostate corresponding to the initially prepared glass, i.e. the reference state, is indicated by the label $O$. Note the general tree-like structures in all four $t$-graphs which appear to be qualitatively similar, despite the different underlying model (atomistic {\em vs.} mesoscopic) and also the different degreee of glass preparation. The color of each vertex indicates the SCC that it belongs to, as we discuss next.     

\subsection{AQS graph topology and strongly connected components (SCCs)\label{subsec:SCC}}

We will probe the topology of the $t$-graphs more deeply by focusing on their SCCs to which any cyclic response must be confined~\cite{Regev2021}, as we explain now. Two mesostates $A$ and $B$ are connected, if on the $t$-graph there is a directed path of $\Up$- and $\Dn$-transitions that leads from $A$ to $B$. Physically, this implies that there is some shearing protocol that, when applied to $A$, gives rise to a deformation pathway terminating in $B$. We say that two mesostates $A$ and $B$ are {\em mutually reachable}, if there is a deformation pathway from $A$ to $B$ as well as one from $B$ to $A$. Mutual reachability is an equivalence relation (in particular, if the pairs $A,B$ and $B,C$ are mutually reachable, so must be the pair $A,C$). Therefore, the vertices of the $t$-graph can be partitioned into equivalence classes under mutual reachability and these classes form its SCCs~\cite{barrat2008dynamical}. Numerical details on how to extract SCCs from $t$-graphs have been provided in Ref.~\cite{Regev2021}.

By construction, transitions between any two mesostates belonging to different SCCs are irreversible: there may be a deformation pathway from one to the other, but not back, since otherwise the pair of states would have been mutually reachable. Thus mutual reachability also partitions the set of transitions between mesostates into reversible ones, i.e. those connecting a pair of mesostates within the same SCC, and irreversible ones, where the two mesostates must belong to different SCCs. Any periodic and hence reversible response to some shear protocol must therefore be confined to a single SCC. The SCCs are thus the ``containers'' of reversible behavior~\cite{Regev2021}. 

\subsection{Comparison of the poorly- and moderately-aged catalogs\label{subsec:PA-WA-catalogs}}

\begin{table}
\caption{
Properties of the $10$ catalogs obtained from poorly-aged (PA) glasses of the mesoscopic model with $N = 32$. The catalogs are labeled by their run number, as given in the first column, while $g_{\rm comp}$ identifies the generation upto which all outgoing mesostate transitions have been identified. The number of mesostates and SCCs found in the catalog are given by $N_{0}$ and $N_{\rm SCC}$, respectively. The last four columns provide statistics about limit-cycles under cyclic shear contained in the catalog (refer to text for details). The last row is a cumulative total over the entries in the corresponding columns. 
}
\rowcolors{2}{white}{black!20}
\begin{tabular}{ *8l }  \toprule \hline
Run  &  $g_{\rm comp}$  & $N_{0}$ & $N_{\rm SCC}$ & $n_{\rm cycles}$ & $N^{\rm supp}_{\rm SCC}$ & $s^{\rm max}_{\rm supp SCC}$ & $n_{\rm cycles}^{\rm max SCC}$ \\ \midrule \hline 
1 & 35 & 26093 & 5817 & 21631 & 4598 & 91 & 97 \\ 
2 & 35 & 59281 & 11084 & 44902 & 8579 & 175 & 84 \\ 
3 & 35 & 28418 & 5963 & 23956 & 4261 & 128 & 116 \\ 
4 & 35 & 131100 & 29478 & 123341 & 24215 & 106 & 104 \\ 
5 & 35 & 48832 & 10374 & 52900 & 9474 & 73 & 67 \\ 
6 & 35 & 89710 & 22955 & 101298 & 21130 & 132 & 116 \\ 
7 & 35 & 46049 & 11498 & 36801 & 9301 & 139 & 124 \\ 
8 & 35 & 145281 & 43409 & 133984 & 34033 & 104 & 67 \\ 
9 & 35 & 52641 & 12854 & 56017 & 11595 & 148 & 124 \\ 
10 & 35 & 49355 & 10155 & 47003 & 7377 & 115 & 153 \\ 
 \midrule \hline 
{\bf ALL} &  n/a  & 676760 & 163587 & 641833 & 134563 &  n/a  & 1052 \\ \bottomrule
\hline
\end{tabular}
\label{tab:N32PAcatalogs}
\end{table}

Tables \ref{tab:N32PAcatalogs} and \ref{tab:N32WAcatalogs} show the properties of the $10$ catalogs with $N = 32$ which were obtained by taking the moderately- and poorly-aged mesoscopic glasses as reference states. The second column lists the number of generations $g_{\rm comp}$ up to which all  outgoing mesostate transitions were identified. Thus $g_{\rm comp} = 39$ means that we have identified every mesostate that can be reached from the reference configuration by a sequence of $39$ $\Up-$ and $\Dn$-transitions. Next, $N_{0}$ and $N_{\rm SCC}$ list the number of mesostates and SCCs contained in the catalog. The last row of each table provides the cumulative totals. We will discuss the results shown in the last four columns later in this section.

\begin{figure}
\includegraphics[width = \columnwidth]{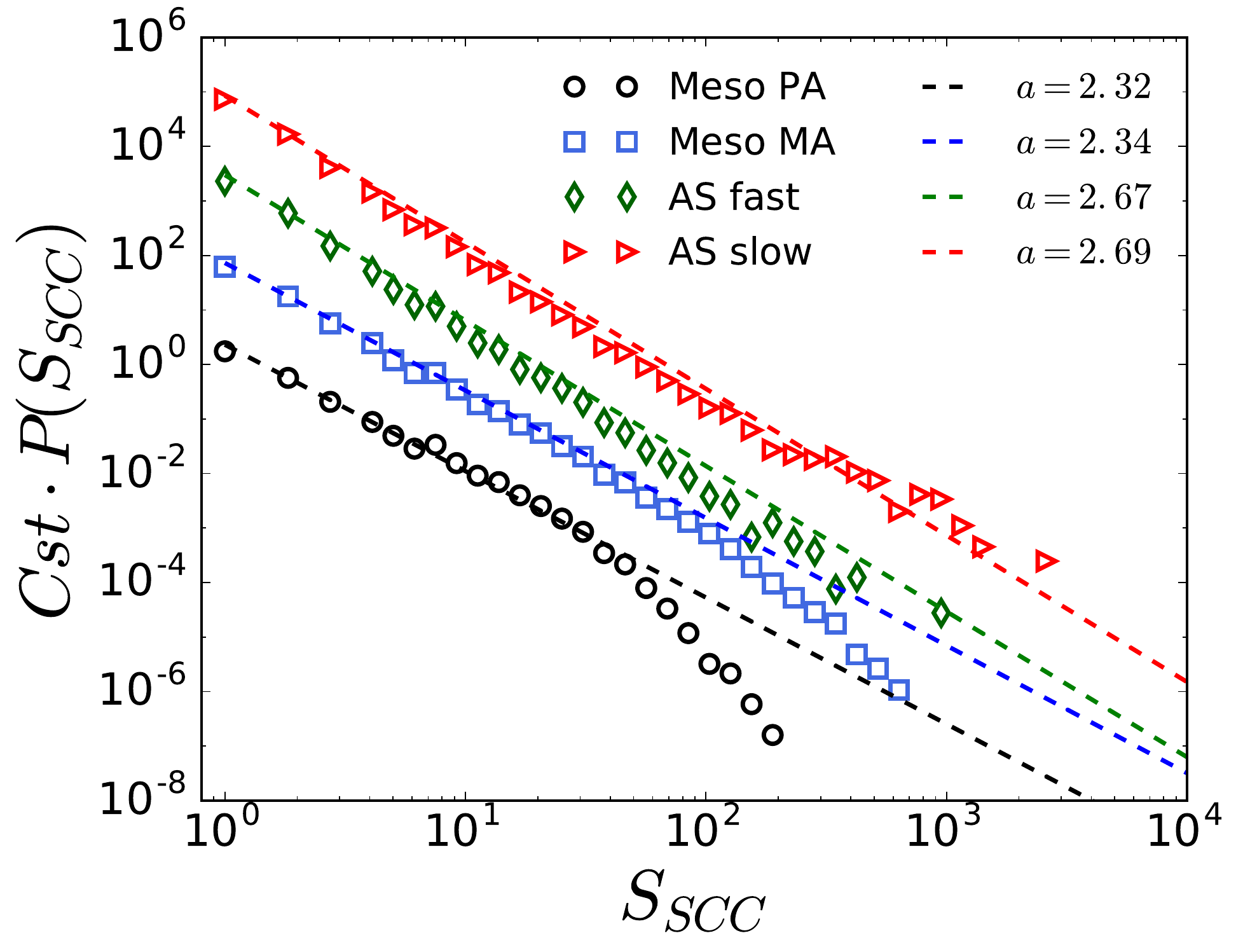}
\caption{\label{fig:scc} Statistics of SCCs vs thermal history \--- Comparison of the SCC size distributions obtained from simulations of the atomistic (AS) and $N = 32$ mesoscopic models (Meso) and distinguished by the extent of aging they have been subjected to: moderately-aged, labeled as Meso MA and AS slow, and poorly-aged, labeled as Meso PA and AS fast, respectively. The dashed lines are power-law fits to the data, which were obtained using a common lower SCC size cut-off of $s_{\rm SCC} = 4$. Curves have been vertically offset for clarity. 
}
\end{figure}

{\it SCC size distributions} \--- 
In Fig.~\ref{fig:scc} we compare the size distribution of the SCCs found in these catalogs. The blue boxes and black circles show the size distribution of SCCs extracted from all $10$ catalogs of the $N=32$ mesoscopic glasses.
All curves have been vertically offset for clarity.  Observe that the size distributions are broad and that the moderately-aged catalogs contain larger SCCs.
Nevertheless, power-law fits using the method of Clauset et al. \cite{clauset2009power} yield a comparable power-law exponent of about $2.3 \pm 0.3$ for both distributions\footnote{For the estimate of the exponent, we considered only SCCs with  sizes  $s_{\rm SCC} \ge  4$, as in earlier work \cite{Regev2021}, where this choice was justified by the empirical observation that small SCCs containing mesostates that were added to the catalog at the last generations are more likely to increase in size, if the catalog is augmented by going to a higher number of generations.}. 
For comparison purposes, we also show the SCC size distributions obtained from our atomistic simulations under slow and fast quench, labeled as AS slow (triangles) and AS fast (diamonds), corresponding to moderately- and poorly-aged glasses. These catalogs reveal similarly broad distributions, with the moderately-aged catalogs containing again larger SCCs, while the fitted power-law exponents $2.7 \pm 0.3$ are comparable. 

We note, however, the presence of a finite-size cut-off around SCC sizes of about $30$ and $100$ for the mesoscopic PA and MA catalogs, respectively. The SCC size distributions obtained from the atomistic simulations do not feature such a cut-off. Although our goal is not to quantitatively map the elastoplastic model onto atomistic simulations, we must ensure that the disordered landscape statistics are comparable between the two types of models. From this point of view, being able to estimate the number of simulated elements of the mesoscopic model, i.e. the system's number of degrees of freedom, is essential for a reasonable comparison which takes also into account possible finite-size effects.

To estimate the equivalent number of simulated elements, one must first  determine the element size of the  elastoplastic model  below which the mechanical description is unresolved. This size corresponds to an upper limit of the characteristic plastic rearrangement size. Several experimental approaches have been performed to estimate the size of rearrangements ranging from direct observations in colloidal systems~\cite{schall_structural_2007} to indirect estimations from strain rate sensitivity analysis in metallic glasses~\cite{ma_nanoindentation_2015}. In all of these cases, the results show that plastic rearrangement cores contain a few dozen particles, so that the overall sizes of these cores range from about two to three particle diameters.

\begin{table}[t!]
\caption{Properties of the $10$ catalogs obtained from moderately-aged (MA) glasses of the $N = 32$ mesoscopic model. Refer to the caption of Table \ref{tab:N32PAcatalogs} for the description of the columns.}
\rowcolors{2}{white}{black!20}
\begin{tabular}{ *8l } \toprule \hline
Run  &  $g_{\rm comp}$  & $N_{0}$ & $N_{\rm SCC}$ & $n_{\rm cycles}$ & $N^{\rm supp}_{\rm SCC}$ & $s^{\rm max}_{\rm supp SCC}$ & $n_{\rm cycles}^{\rm max SCC}$ \\ \midrule \hline 
1 & 39 & 46059 & 8148 & 3510 & 857 & 269 & 7 \\ 
2 & 39 & 36279 & 8164 & 1732 & 363 & 451 & 11 \\ 
3 & 39 & 130733 & 33324 & 3933 & 1148 & 542 & 129 \\ 
4 & 39 & 19344 & 4244 & 1659 & 490 & 207 & 3 \\ 
5 & 39 & 147476 & 49335 & 989 & 437 & 133 & 2 \\ 
6 & 39 & 64096 & 11678 & 1731 & 643 & 166 & 2 \\ 
7 & 39 & 117680 & 30721 & 6189 & 1809 & 244 & 58 \\ 
8 & 39 & 64693 & 12657 & 5317 & 1219 & 651 & 179 \\ 
9 & 39 & 118964 & 33857 & 3067 & 1143 & 141 & 12 \\ 
10 & 39 & 91758 & 26814 & 8516 & 2011 & 201 & 127 \\ 
 \midrule \hline 
{\bf ALL} &  n/a  & 837082 & 218942 & 36643 & 10120 &  n/a  & 530 \\ \bottomrule
\hline
\end{tabular}
\label{tab:N32WAcatalogs}
\end{table}

The determination of this length scale in atomistic simulations poses  several difficulties. First, the presence of avalanches makes it challenging to identify the individual rearrangements. Second, there is no  method yet to spatially distinguish between the non-linear and non-affine elastically strained zones from the non-reversible plastic responses. Finally, another  complication arises from the fact that the same zone can contain several slip directions in a realistic particle system~\cite{nicolas_orientation_2018,barbot_local_2018}, resulting in an effective higher density of potential rearrangements than that of a scalar description. Several approaches have been implemented to deal with these difficulties. They rely on the analysis of long-range elastic  fields~\cite{albaret_mapping_2016,nicolas_orientation_2018}, the quantitative calibration of elasto-plastic models~\cite{Patinet-CRP21}, the calculation of the spatial extension of rearrangements~\cite{puosi_probing_2015}, the strain's spatial correlations~\cite{lemaitre_rate-dependent_2009,nicolas_spatiotemporal_2014}, and the reproduction of the mechanical response from the spatial density of barriers~\cite{Patinet-PRL2020-Bauschinger}. These approaches, particularly those using a two-dimensional system under AQS loading like ours, lead to a consistent estimate of the linear size of plastic rearrangements lying between $3$ and $7$ particle diameters. For our atomistic system containing $1024$ atoms, these bounds lead to an equivalent   mesoscopic system  size between $N=5$ and $10$. Indeed, simulations of the mesoscopic model performed for system sizes $N=8$ (not shown here) produce an SCC size distribution where a finite size cut-off is absent, but the scaling exponent is comparable.  We thus ascribe the absence of a finite size cut-off in the SCC size distributions of the atomistic simulations  to a finite size effect.

Note that while the $t$-graphs and SCC size distributions obtained from our mesoscopic and atomistic model are qualitatively similar, the dependence of these on the degree of aging is rather weak. In other words, the topology of the $t$-graphs alone does not appear to contain features that are directly linked to the different amount of aging these samples have been subjected to. As we will show next, the effect of aging on the samples reveals itself when we combine the topological features of the $t$-graphs with additional physical properties, such as the prevalence of cycles, the plastic strain and the strain stability ranges associated with the mesostates and their SCCs.   

{\it Prevalence of cycles} \--- We next turn to the population of cycles in our catalogs. We are again interested in cycles that can be traversed under a {\em symmetric} cyclic shear protocol: $0 \to \epsilon_{\rm amp}  \to  -\epsilon_{\rm amp} \to 0$ with some shear amplitude $\epsilon_{\rm amp}$. We consider every mesostate in our catalog that is stable at zero strain and apply this cyclic shear protocol, checking whether a cyclic response has set in or not. The column labeled $n_{\rm cycles}$ of Tables \ref{tab:N32PAcatalogs} and \ref{tab:N32WAcatalogs} lists the total number of distinct cycles found in our catalogs obtained from our moderately- and poorly-aged mesoscopic glasses. We find that the poorly-aged catalogs contain a significantly  larger number of cycles, although the total number of mesostates in these catalogs is comparable ($836082$ 
and $676760$ mesostates, respectively). 

As we have noted before, the mesostates forming a cyclic response must all be confined to a single SCC, i.e. a cycle cannot span multiple SCCs.  We therefore ask next how the cycles found in the catalogs are distributed across the available SCCs. In particular, we ask for the number of SCCs that support at least one symmetric cycle,  
which we define as $N^{\rm supp}_{\rm SCC}$ and list in Tables \ref{tab:N32PAcatalogs} and \ref{tab:N32WAcatalogs}. 
For ease of comparison, we have put together in Table \ref{tab:CatalogCyclesComp} the cumulative totals  listed 
in the last lines of these tables along with the corresponding data from our atomistic simulations.

\begin{table}[h!]
\caption{
Comparison of the cumulative totals of the number of mesostates $N_0$, SCCs $N_{\rm SCC}$, and SCCs that support symmetric cycles $N^{\rm supp}_{\rm SCC}$. The top two rows show data  for the poorly-aged (PA) and moderately-aged (MA) mesoscopic glasses. The bottom two rows compare these quantities for the fast and slow cooled atomistic glasses.  Refer to text for further details and the Tables \ref{tab:glv2catalogs} and \ref{tab:waglv2catalogs} in Appendix \ref{sec:catalog-properties} for the sample-by-sample characterization of the atomistic catalogs. 
}
\centering
\rowcolors{2}{white}{black!20}
\begin{tabular}{ *4l } \toprule \hline
Catalogs   & $N_{0}$ & $N_{\rm SCC}$ & $N^{\rm supp}_{\rm SCC}$ \\ \midrule \hline 
{\bf Meso PA}  & 676760 & 163587  & 134563 \\ 
{\bf Meso MA}  & 837082 & 218942  & 10120 \\
\midrule \hline
{\bf AS Fast}  & 459508 & 210864  & 10933 \\ 
{\bf AS Slow}  & 555332 & 244334  & 5863   \\ 
\bottomrule
\hline
\end{tabular}
\label{tab:CatalogCyclesComp}
\end{table}

Starting with the mesoscopic glasses, there is again  a stark contrast between catalogs obtained from poorly-aged (PA) and moderately-aged (MA) samples (first two rows of Table \ref{tab:CatalogCyclesComp}). 
In the MA glasses the symmetric cycles are contained in a relatively small fraction of SCCs ($10120$  out of a total of $218942$ available ones), while for the poorly-aged catalogs a large fraction of SCCs supports at least one such cycle ($134563$ SCCs that support symmetric cycles out of a total of $163587$). From Tables \ref{tab:N32PAcatalogs} and \ref{tab:N32WAcatalogs}, we see that this is true also for the individual catalogs.  It is thus apparent that in the moderately-aged catalogs a relatively small fraction of SCCs support most of the cycles found, while in the poorly-aged catalogs the opposite is the case and almost every SCC supports at least one cycle. A similar, albeit less pronounced behavior is seen also in our atomistic simulations, {\em cf.} the last two rows of Table \ref{tab:CatalogCyclesComp}. Note that the cumulative data for poorly-aged (moderately-aged) initial states have been sampled from $8$ $(30)$ catalogs (Tables \ref{tab:glv2catalogs} and \ref{tab:waglv2catalogs} in Appendix \ref{sec:catalog-properties}), so that it is hard to compare the overall number of cycles.  Nevertheless, we observe also in our atomistic simulations that the number of cycle supporting SCCs in the poorly-annealed catalogs appears to be disproportionally larger.  

We finally consider the largest SCCs that support symmetric cycles, comparing their sizes $s_{\rm suppSCC}^{\rm max}$ and the number of cycles they contain $n_{\rm cycles}^{\rm max SCC}$.
These numbers are shown in the last column of Tables \ref{tab:N32PAcatalogs} and \ref{tab:N32WAcatalogs}. Again, we find contrasting behavior. The largest cycle supporting SCCs found in the moderately-aged catalogs are generally larger than those in the poorly-aged ones, but despite of this,  they contain fewer cycles. 

\subsection{The disorder landscape}

Our results for the prevalence of symmetric cycles
can be summarized as follows: while the poorly-aged catalogs contain a large number of such cycles which are distributed across a large number of SCCs of various sizes, we find that the opposite is true for the catalogs obtained from the well-aged samples. For the latter, the number of symmetric cycles contained is far less and these cycles are confined to a small subset of available SCCs. 

\begin{figure*}
\includegraphics[width=\textwidth]{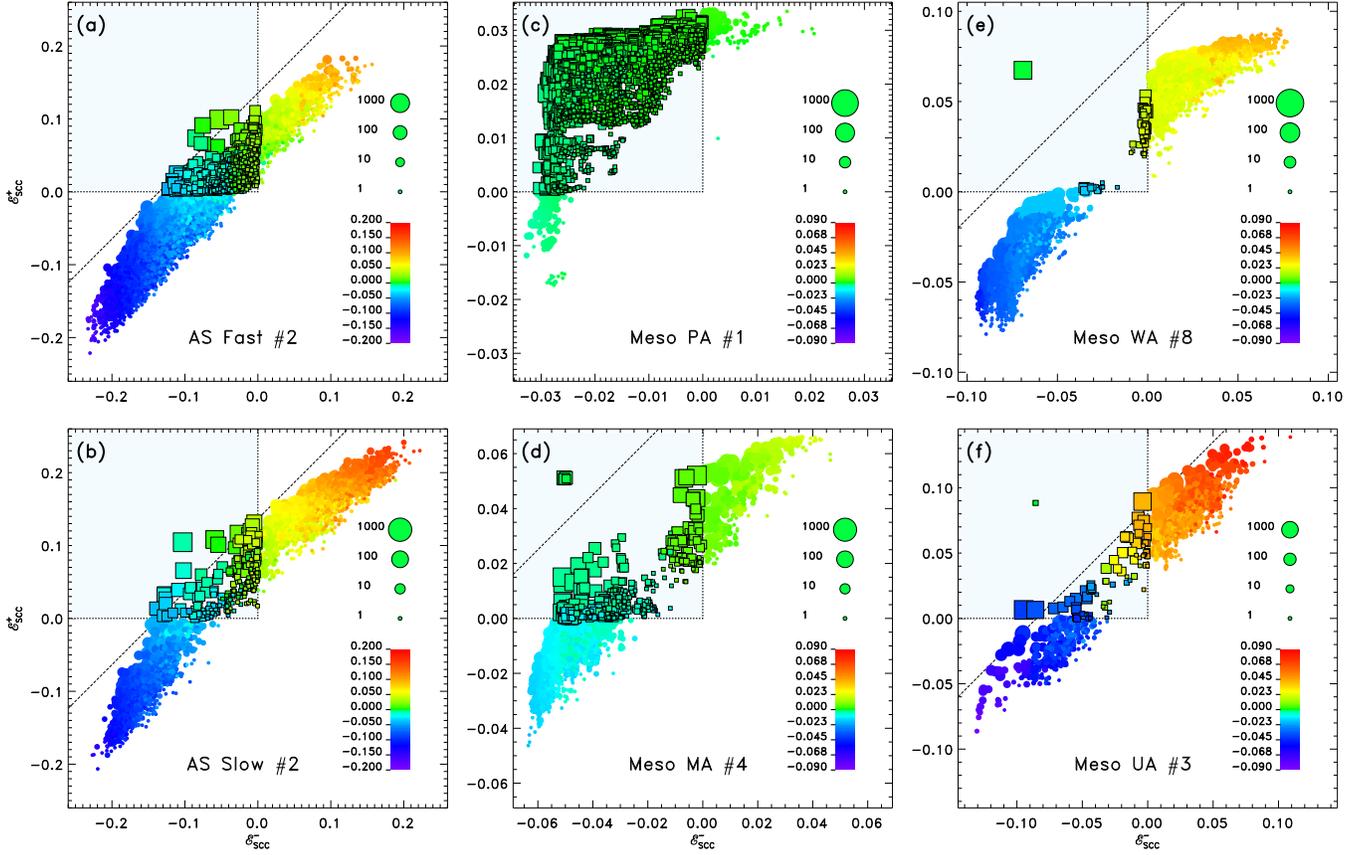}
\caption{\label{fig:scc-scatter} 
The coarse-grained disorder-landscape, atomistic {\em vs.} mesoscopic models, and the effect of aging -- The four panels (a) -- (d) show the scatter plots of the SCCs found in catalogs obtained from atomistic (first column) and our $N = 32$ mesoscopic simulations (second column). Panels (e) and (f) depict the disorder landscape extracted from increasingly better-aged samples of the mesoscopic model. Each symbol represents an SCC, while the size and color correspond to the size of the SCC and the average plastic strain $\epsilon^{\rm pl}$ of the mesostates constituting that SCC, as indicated in the legends. Each SCC has at least one  $\Up$- and one $\Dn$-transition that leads to another SCC,  
and we denote by    $\mathcal{E}^\pm_{\rm SCC}$ the threshold strains to trigger these transitions. As explained in the text, taking the extremes of these exit strains, the corresponding interval $(\mathcal{E}^-_{\rm SCC}, \mathcal{E}^+_{\rm SCC})$ provides a range of strain values over which the system will be trapped in that SCC. These strains are used as coordinates for placing the SCC in the plot. Box-shaped symbols indicate that the SCC supports at least one cycle under symmetric cyclic shearing. The diagonal dashed lines corresponds to the average SSC strain range of Eq.~\ref{eq:SCC_range}, estimated as $ \mathcal{E}^+_{\rm SCC} - \mathcal{E}^-_{\rm SCC} = \Sigma_{\rm ss}/\mu$, where  $\Sigma_{\rm ss}$ is the steady-state stress under monotonous strain loading. Refer to text for further details.   
} 
\end{figure*}

In order to understand better the difference of the disorder landscape arising from well-aged and poorly-aged samples, we coarse-grain the $t$-graph to the level of SCCs, since -- as we have shown -- any cyclic response must be confined to a single SCC. Every SCC has at least one outgoing $\Up$- and one outgoing $\Dn$-transition. Let us denote the states from which these outgoing transitions originate as the $\Up$- and $\Dn$-exits of the SCC. 
Suppose now that the SCC has only one $\Up$- and one $\Dn$-exit  
and denote the threshold strains triggering these exiting transitions as $\mathcal{E}^\pm_{\rm SCC}$. Consequently, given any mesostate $A$ belonging to that SCC and applying strains confined to the interval 
$\mathcal{E}^-_{\rm SCC} < \epsilon < \mathcal{E}^+_{\rm SCC}$, the resulting sequence of mesostates must remain confined to the SCC. This follows from the observation made before, namely that for any mesostate $A$,  $\epsilon^+[A] < \epsilon^+[\Up A]$ and $\epsilon^-[\Dn A] < \epsilon^-[A]$. 

In the case of multiple $\Up$- or $\Dn$-exits from an SCC, we define $\mathcal{E}^+_{\rm SCC}$ and $\mathcal{E}^-_{\rm SCC}$ as the largest,  respectively lowest, strain triggering the outgoing transitions. It actually turns out that for the SCCs considered in our catalogs only a very small fraction of SCCs have multiple $\Up$- or $\Dn$-exits\footnote{For the $6$ catalogs shown in Fig.~\ref{fig:scc-scatter} the percentages of SCCs with more than one $\Up$- or $\Dn$-exits are: $1.7\%$ (AS Fast), $1.7\%$ (AS Slow),  $2.9\%$ (PA), $3.6\%$ (MA), $0.9\%$ (WA), and $7.2\%$ (UA). The low number of exits from SCCs is also apparent from the transition graph excerpts shown in Figs.~\ref{fig:tgraphs} and \ref{fig:tgraphs_wa}.
}
Assuming therefore that each SCC has exactly one outgoing $\Up$- and $\Dn$-transition, it follows that in order for  the SCC to support cyclic response under the strain protocol $0 \to \epsilon_{\rm amp} \to -\epsilon_{\rm amp} \to 0 \cdots$, we must require that 
$\mathcal{E}^+_{\rm SCC} > \epsilon_{\rm amp}$ and $\mathcal{E}^-_{\rm SCC} < -\epsilon_{\rm amp}$. 
In particular, this implies that 
\begin{equation}
    \mathcal{E}^-_{\rm SCC} < 0 < \mathcal{E}^+_{\rm SCC}.
    \label{eqn:cycle-quadrant}
\end{equation}

Distinguishing the SCCs by (i) their size, and (ii) whether they support a symmetric cycle or not, we now ask how these SCCs are scattered in the plane spanned by $\mathcal{E}^-_{\rm SCC}$ and $\mathcal{E}^+_{\rm SCC}$. Panels (a) and (b) of Fig.~\ref{fig:scc-scatter} show the 
SCC scatter plots obtained from single catalogs of our atomistic poorly-aged and  moderately-aged samples, while panels (c) and (d) show the same for catalogs obtained from our mesoscopic poorly-aged and moderately-aged $N = 32$ samples. Panels (e) and (f) show SCC scatter plots obtained from even further aged mesoscopic samples, with an average of $150$ and $4000$ aging steps per site, respectively (details of these catalogs are provided in Appendix \ref{sec:catalog-properties}). 
In each panel of the figure the number after the $\#$ sign indicates the particular sample from which the data shown came from.
The size of the symbols represent the size of the SCCs, as indicated in the legend, while the boxed symbol shape indicates that the SCC actually supports a limit-cycle, as determined by inspecting our catalogs. The highlighted upper left quadrant of each plot corresponds to the region where the inequality  \eqref{eqn:cycle-quadrant} holds.  Since this is the region where any SCC which supports cyclic response under symmetric oscillatory shear must be located, we will refer to it as the {\em cycle-quadrant}. 

We start with a comparison of the poorly-aged (PA) and moderately-aged (MA) SCC scatter plots obtained from our atomistic and mesoscopic glasses, panels (a) -- (d). 
Comparing the catalogs obtained from the PA samples, panels (a) and (c), with those of the MA samples, panel (b) and (d), we see that in all cases the cycle supporting SCCs (boxes) are indeed confined to the cycle-quadrant, i.e. the highlighted region in the top left part of the figure, as they should. 
Moreover, note the relative sparsity of cycle-supporting SCCs in the atomistic (b) and mesoscopic (d) MA samples, when compared with their poorly-aged counterparts, panels (a) and (c). This is consistent with our earlier observation, namely that relative to the poorly-aged samples, in the MA catalogs only a small fraction of SCCs actually support symmetric cycles.

Plotting the SCCs against their exit strains $(\mathcal{E}^-_{\rm SCC},\mathcal{E}^+_{\rm SCC})$ also visualizes possible correlations in the locations of cycle supporting SCCs. For the poorly-aged samples, panels (a) and (c), these SCCs fill out the cycle-quadrant rather uniformly and the extent to which this region is filled seems to be limited mainly by the size of the catalog we have sampled, i.e. the number of generations we tracked\footnote{Note the different strain ranges when comparing  atomistic and mesoscopic catalogs, panels (a) -- (d) of Fig.~\ref{fig:scc-scatter}. While the atomistic catalogs sample SCCs whose exit strains well exceed the yield-strain (about $0.13$), the mesoscopic catalogs stay well below yielding. This is due to the fact, that the $N = 32$ mesoscopic systems correspond to much larger atomistic systems, then the ones we simulated.}.  
This is in contrast to the case of the moderately-aged samples, panels (b) and (d): not only are there fewer SCCs in the cycle-quadrant, but these SCCs tend to cluster around its boundaries, $\mathcal{E}^+_{\rm SCC} = 0$ and $\mathcal{E}^-_{\rm SCC} = 0$, implying thereby that these SCCs can only support cycles with low amplitudes of a symmetrical shear protocol.  In fact, for the mesoscopic samples we find that the scarcity of SCCs within the cycle-quadrant and their clustering near its boundary becomes even more pronounced when the samples are aged more, as shown in the SCC scatter plots of panels (e) and (f) which were generated from samples that underwent $150$ and $4000$ aging steps per site, respectively. 

Thus panels (a) -- (d) reveal that the SCC scatter plots obtained from our mesoscopic model are qualitatively very similar to their atomistic counterparts: our mesoscopic model captures rather well the difference of the samples due to their aging as well the spatial distribution of the SCCs in the plane plane of exit strains $(\mathcal{E}^-_{\rm SCC},\mathcal{E}^+_{\rm SCC})$. 

{Before proceeding, we should note that there are sample-to-sample fluctuations in the scatter plots obtained from the individual glasses. This is also apparent in the variation of catalog properties listed in Tables \ref{tab:N32PAcatalogs} and \ref{tab:N32WAcatalogs}, as well as in the tables for the other catalogs given in Appendix \ref{sec:catalog-properties}. In particular, the spatial population of SCCs in the cycle-quadrant varies from sample to sample. Moreover, within a given sample the populations of SCCs in the 
$(\mathcal{E}^-_{\rm SCC},\mathcal{E}^+_{\rm SCC})$-plane does not perfectly display the statistical 
$\mathcal{E}^\pm_{\rm SCC} \to - \mathcal{E}^\mp_{\rm SCC}$ symmetry which arises under interchange of the forward and reverse shearing directions, even though the number of SCCs shown in these plots are rather large. Nevertheless, the features we have been discussing so far and in the following are typical and appear to be robust from sample to sample.}

Having established that the SCC scatter plots are a good proxy to probe toplogical features of the disorder landscape, we next look more closely at the effect of aging on our mesoscopic glasses. 
Panels (c) -- (f) of Fig.~\ref{fig:scc-scatter} show SCC scatter plots obtained from increasingly better aged samples of our mesoscopic glass, which apart from the PA, MA and WA samples we considered so far includes now also an ultra-aged (UA) glass, obtained from a treatment with $4000$ aging steps per site. 

Note that the moderately-aged (MA), well-aged (WA), and ultra-aged (UA)  samples each display distinct outlier SCCs in the cyclic quadrant. For the MA sample these SCCs are located around $(\mathcal{E}^-_{\rm SCC},\mathcal{E}^+_{\rm SCC}) = (-0.05,0.05)$, while for the WA samples these are found at larger strains. These SCCs turn out to be formed by mesostates that can be reached from the initially prepared glass by strain deformation protocols that do not go beyond the stress-peak and hence do not suffer the subsequent large stress-drop. 

To understand why with increased aging the cyclic quadrant becomes less densely populated by SCCs and why these tend to cluster near its boundaries, we consider next the plastic strains. Recall that with each mesostate $A$ we associate an elastic branch in the stress-strain plane. In the case of our mesoscopic model, this branch is by construction linear and the plastic strain  $\epsilon^{\rm pl}[A]$ associated with the branch is the (extrapolated) value of the strain where the stress vanishes. By averaging over the plastic strains of the mesostates that belong to an SCC, we obtain a coarse-grained plastic strain for each SCC. The colors of the plot symbols shown in Fig.~\ref{fig:scc-scatter} represent the plastic strains of the SCCs, as indicated by the color table legends. Note that for the mesoscopic samples, panels (c) -- (f), we have color-coded the same range of plastic strains. Thus the shift of colors towards blue and red as the samples get better aged indicates that the magnitudes of typical plastic strains increase with aging.

Moreover, we see that the distribution of plastic strains across SCCs is strikingly different for the differently aged samples. The well- and ultra-aged samples reveal a clear bi-modal distribution of plastic strains, characterized by very few SCCs that have vanishing plastic strains\footnote{In all four panels of Fig.~\ref{fig:scc-scatter} the green color used in the legend of the SCCs sizes corresponds to the color-coding of vanishing plastic strains.}. For the poorly-aged and medium-aged samples, panels (c) and (d), the distribution of plastic strains appears to be unimodal, with a large number of SCCs, particularly those in the cycle quadrant, having plastic strains of very small magnitude. 

The bi-modal nature of the SCC plastic strain distribution for the  WA and UA samples is a direct consequence of the fact that there is a large  stress-drop right after the stress peak has been reached under uniform applied shear, {\em cf.} Fig.~\ref{fig:monotonous-loading}. In fact, we verified that for the WA and UA samples all SCCs (except the small outlier SCCs) are reached from the initial glass configuration by deformation pathways that experience the stress drop, which then in turn produces a corresponding jump in the plastic strain (see also the $t$-graph excerpts shown in  Fig.~\ref{fig:tgraphs_wa} of Appendix \ref{sec:catalog-properties}, where transitions accompanied by larger stress drops have been marked). The better the aging, the larger the stress drops, and hence the larger the jumps in plastic strain. 
Moreover, these gains in plastic strain due to the experienced stress drop are apparently very hard to undo by subsequently shearing in the reverse direction. We find that under shearing in the  reverse direction the sample has now been significantly softened, i.e. it has become more plastic, indicating a rejuvenation of the sample~\cite{barbot2020rejuvenation}.  
Thus for the well-aged  and ultra-well-aged samples the diagonal $\mathcal{E}^-_{\rm SCC} + \mathcal{E}^+_{\rm SCC} = 0 $ divides the plane of exit strains into an upper and lower half. SCCs located in the upper (lower) half of the plot are SCCs whose mesostates were reached by passing through the forward (reverse) stress peak. This is also consistent with the excerpts from the corresponding transition graphs shown in Fig.~\ref{fig:tgraphs_wa} of Appendix \ref{sec:catalog-properties}.  

As our cyclic shear simulations show, even for the WA and UA samples, and strain amplitudes close to but below the irreversibility transition, cyclic response may eventually be attained, but after a long transient.
In particular, for the  WA and UA samples we find that with increasing system size the transition to irreversibility becomes abrupt, meaning that either we reach cyclic response after a few driving cycles (typically $1$ or $2$ cycles) or not at all, implying a rather sharp and possibly discontinuous transition from reversibility to irreversibility. Our simulations indicate that this transition becomes smoother when keeping the system size fixed and the samples are less aged, or when at fixed aging steps per site, we reduce the system size. Thus for example for system sizes $N = 16$ and $N=32$, the well-aged samples are able to attain limit-cycles rather quickly and at strain amplitudes that are well above the onset of yielding under uniform shear, marked by the location of the stress peak.   The  amplitudes beyond which the probability of finding a cycle is less than $2\%$ have been marked by the triangles in the monotonous loading curves of Fig.~\ref{fig:monotonous-loading}. They are located beyond the stress peak. In fact, these observations are consistent with findings in recent work by one of us on periodically sheared $3d$ atomistic glass formers~\cite{adhikari2022yielding}. There it was found that small samples that were moderately- or well-aged exhibit cyclic response at amplitudes well beyond the value of the strain at the stress peak. As the size of the samples increases, a sharp irreversibility transition at the stress peak is recovered. We should note however, that in Ref.~\cite{adhikari2022yielding} such behavior was found to be the case only for totally {\em asymmetric} shear protocols of the form $0 \to \epsilon_{\rm amp} \to 0 \to \epsilon_{\rm amp} \cdots$.   

We conclude with a discussion of the spatial arrangement of SCCs along a strip-like region in the $(\mathcal{E}^-_{\rm SCC},\mathcal{E}^+_{\rm SCC})$, that is clearly evident for the atomistic systems as well as the  WA and UA mesoscopic  samples in the SCC scatter plots of Fig.~\ref{fig:scc-scatter}. The diagonal dashed line in each of the plots corresponds to
\begin{equation}
\label{eq:SCC_range}
 \Delta \mathcal{E}_{\rm SCC} = \mathcal{E}^+_{\rm SCC} - \mathcal{E}^-_{\rm SCC}.     
\end{equation}
As one would expect, the larger the strain range $\Delta \mathcal{E}_{\rm SCC}$ over which mesostates are trapped within an SCC, the larger the size of the SCC itself. This trend is clearly visible in all six  panels of the plots. The smallest (and most numerous) SCCs are clustered around small values $\Delta \mathcal{E}_{\rm SCC}$, while particularly for the WA and UA samples there appears to be a value $\Delta \mathcal{E}_{\rm SCC} = \Delta \mathcal{E}_{\rm max}$ beyond which it is unlikely to find SCCs, except for the outlier SCCs that we have associated with mesostates not having experienced the stress-peak. A naive estimate for $\Delta \mathcal{E}_{\rm max}$ can be made as follows. Denote by $\Sigma_{\rm ss}$ the steady-state yield stress reached under monotonous loading ({\em cf.} Fig.~\ref{fig:monotonous-loading}). Assuming that between $-\Sigma_{\rm ss}$ and $\Sigma_{\rm ss}$ the system responds purely elastically,
we obtain the  estimate $\Delta \mathcal{E}_{\rm max} = \Sigma_{\rm ss}/\mu$. From Fig.~\ref{fig:monotonous-loading} we find that for the mesoscopic samples  $\Sigma_{\rm ss} \approx 0.85$, 
while for the atomistic samples $\Sigma_{\rm ss} \approx 2.4$. The dashed lines shown in Fig.~\ref{fig:scc-scatter} correspond to these choices, i.e. $\mathcal{E}^+_{\rm SCC} - \mathcal{E}^-_{\rm SCC} = \Sigma_{\rm ss}/\mu$. 

\section{Discussion}

We have introduced a depinning-like mesoscopic model of amorphous plasticity characterized by a tunable aging and embedded in a quenched disorder landscape. When driven by an externally applied shear, the model recovers many phenomena exhibited by sheared amorphous solids: a brittle-to-ductile transition under monotonous strain loading, as well as an irreversibility transition under symmetric oscillatory shear, i.e. of the form $0 \to \epsilon_{\rm amp} \to 0 \to - \epsilon_{\rm amp} \to 0$, and its dependence on the extent the sample has been aged. 

We find that the irreversibility transition exhibits a strong dependence on system size as well as on the extent of prior aging of our mesoscopic glasses. Close to the transition, poorly-aged systems show a power-law behavior for both the duration of the transient (number of loading cycles needed to reach the limit cycle) and the mean period of the cyclic response (measured in units of the number of driving cycles). Moreover, with increasing system size, the strain value at which the irreversibility transition occurs seems to converge to a well defined value in the infinite system size limit. 

In the case of the better aged samples, we find that cyclic response under oscillatory shear emerges after  only a few loading cycles. The dependence on system size is more pronounced in this case. Samples of small size exhibit a cyclic response containing many plastic events and this response continues up to strain amplitudes at which the system would have yielded under monotonous shear loading. However, as the sample size increases, the cyclic response becomes more elastic and the range of strain amplitudes at which it is exhibited shrinks. Changing the system size in our moderately- and well-aged samples allows us to gradually transition from a cyclic response whose phenomenology is characteristic of poorly-aged glasses to one where this cyclic response is dominantly elastic at larger sizes.

In order to better understand the dependence of the dynamics of our mesoscopic model on the prior aging, we turned next to the study of the transition graphs ($t$-graphs) which capture the transitions between accessible elastic branches via plastic events. The topology of the $t$-graphs encodes the dynamics under arbitrary shear loading protocols and thus provides a complementary tool to characterize the disorder landscapes underlying our differently aged systems. We considered a particular topological quantity characterizing the $t$-graphs, its strongly connected components (SCCs), since any cyclic response has to be confined to a single SCCs. The size distribution of SCCs sampled from both atomistic and mesoscopic simulations of differently aged samples all follow a power-law with an exponent that varies little with the extent of aging but is slightly  smaller for the mesoscopic systems than for the atomistic ones. 

A closer inspection that also takes into account physical properties associated with the SCCs, in particular their range of stability and typical plastic strains, turns out to be extremely informative. We find that the sample age induces a gradual phase separation between domains of stability centered either on the initial state or at a finite positive or negative plastic strain. The complex age-dependence of the interplay between the amplitude of the center of the domain and the width of the stability ranges has important consequences on the accessibility of limit cycles depending on the particular parameters of the cycling protocols. 

Our findings have also implications for memory formation in amorphous solids. Cyclic response under oscillatory shear can encode information and thus form a “memory”
about the forcing that caused the response~\cite{keim2019memory}. Viewed within the framework of the $t$-graphs any periodic response must be confined to one of its SCCs. Thus the evolution under oscillatory shear is primarily a search for a confining SCC. In fact, such SCCs not only contain the cycle forming the cyclic response, but a hierarchy of nested cycles, one of which forms the cyclic response. A hierarchical organization of cycles is typically associated with return point memory~\cite{munganterzi2018structure,Mungan-PRL19,Regev2021}. In particular, the size of an SCC, i.e. the number of configurations they contain, can be regarded as a proxy for memory capacity~\cite{Regev2021}. 

Since we find that the distribution of SCC sizes is broad, irrespective of the thermal histories of the  glasses from which these distributions were sampled, this suggest a high memory capacity even for well-aged glasses. However, a closer look at the stability ranges of the SCCs found in these glasses, reveals that only the poorly-aged samples have a large abundance of SCCs that can support symmetric cyclic shearing protocols. Contrastingly, in the case of the well-aged glasses very few SCCs support cyclic response to such oscillatory shear protocols. We find that those that do are characterized by loading/driving histories that did not experience the stress-peak and subsequent stress drop. Consequently, their cyclic response is largely elastic and confined to few and relatively small SCCs. 

On the other hand, loading histories in which  a stress peak and subsequent stress drop are encountered, invariably give rise to rejuvenation of the sample which is also accompanied by a jump of the plastic strain to non-zero values. As a result, a large number of SCCs become dynamically accessible. 
However, due to the jumps in plastic strain, these SCCs will only support cyclic response to oscillatory shear if the shear strain is centred around the value of their plastic strain.

Having  demonstrated that our mesoscopic model reproduces key features of amorphous solids under variable athermal quasistatic loading, we conclude with a discussion of possible directions for future research.  
Compared to atomistic models, the computational cost of simulation of  mesoscale models is rather low, allowing us to perform extensive numerical computations as well as probing system sizes not accessible to atomistic simulations. In this context, it would be nice to understand better the complex interplay between finite-size effects and the degree of aging that we have observed under oscillatory shear. In the same vein, a detailed statistical analysis of the spatial structure and correlations of sites that undergo plastic activity will be of interest both near the yielding transition, and also in the evolution of the transients toward cyclic response under oscillatory shear. In this context, it would be relevant to understand how the spatial structure of sites of plastic activity associated with transitions within an SCC correlates with the size of the SCC and its stability range. In fact, one can regard the set of such active sites as a fingerprint of its SCC and ask how this set changes under transitions to neighbouring SCCs, thereby defining an overlap function between SCCs. Since SCCs are containers of periodic response, the strength of such overlaps will have implications for memory formation. Strong overlaps would imply that similar cyclic responses can be realized in neighbouring SCCs. At the same time, such overlaps can also be used to characterize in greater detail the topology of the disorder landscape and its possible hierarchical organization.

\begin{acknowledgments}
This project has received funding from the European Union’s Horizon 2020 research and innovation programme under the Marie Skłodowska-Curie grant agreement No. 754387.
MM would like to thank Monoj Adhikari and Srikanth Sastry for many useful discussions and for sharing their insights on cyclic shearing and the evolution of plastic strain.  
IR and MM acknowledge the kind hospitality of ESCPI which was made possible in part by  Chaire Joliot awards.
The authors would like to thank Lila Sarfati, Ga\"el Tejedor for a careful reading of the manuscript and useful comments. 
MM was funded by the Deutsche Forschungsgemeinschaft (DFG, German Research Foundation) under Projektnummer 398962893, the Deutsche Forschungsgemeinschaft (DFG, German Research Foundation) - Projektnummer 211504053 - SFB 1060, and by the Deutsche Forschungsgemeinschaft (DFG, German Research Foundation) under Germany’s Excellence Strategy - GZ 2047/1, Projekt-ID 390685813. 
MM also acknowledges partial support by the TRA Modelling (University of Bonn) as part of the Excellence Strategy of the federal and state governments.  
IR was funded by the Israel Science Foundation grant no. (ISF) through Grant No. 1301/17.
\end{acknowledgments}

\appendix

\section{Atomistic simulations\label{sec:atomistic-simulations}}
Atomistic simulations were performed on a two-dimensional binary system with $N=1024$ particles of two sizes, where half the particles are $1.4$ times larger than the other half. We used a two-body radially-symmetric interaction introduced in Ref.~\cite{lerner2009locality} and used in Ref.~\cite{regev2013onset}, employing the same units of temperature and time discussed there. The initial sample is prepared by first simulating the system at a high temperature in a liquid state, and then quenching the liquid to zero temperature.
We used two different preparation protocols to obtain soft and hard glasses. To obtain a soft glass, starting from $T=1$ we equilibrated the system for $t=20$ simulation time units and then reduced the  temperature to $T=0.1$ and equilibrated for another $t=50$.
To obtain a hard glass, starting from $T=1$ we cooled the system to $T=0.1$ in steps of $\Delta T=0.025$, where at each step the system was equilibrated for $t=10$. Once an initial solid sample was prepared, it was sheared quasistastically using a standard AQS protocol: at each strain step, the system is sheared using the Lees-Edwards boundary conditions~\cite{lees1972computer} such that the total strain increases by $10^{-4}$. Immediately after strain is applied, the energy is minimized using the FIRE minimization algorithm~\cite{bitzek2006structural}.

\begin{figure}[t!]
\includegraphics[width=0.9\columnwidth]{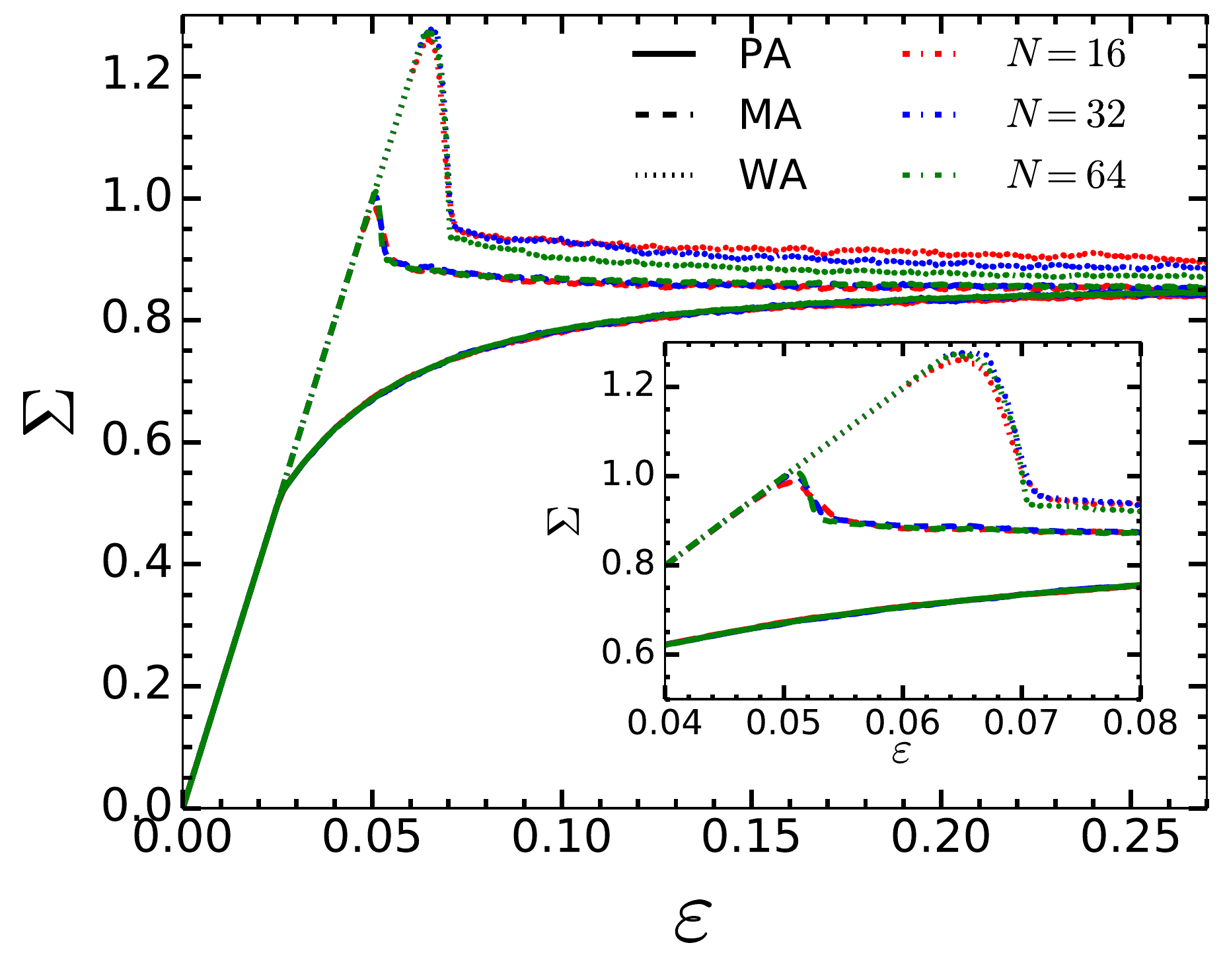}
\caption{ Stress-strain curves upon monotonous loading for various system sizes and thermal histories. The inset shows a blow-up of the region near the stress peak. }
\label{fig:meso_uni_shear}
\end{figure}

\section{Uniform shear of mesoscopic glass\label{sec:meso_uniform_shear}}
Fig.~\ref{fig:meso_uni_shear} shows the dependence of stress under monotonous strain loading on system size and aging. 
Different colors correspond to different system sizes, as indicated in the legend, while the line shapes correspond to the different degrees of aging. 
The curves have been obtained at various extent of aging and for systems of size $N = 16 (1000), \ 32 (500), \ 64 (250)$, where the numbers in parenthesis indicate the number of realizations used to obtain our results.
While the poorly-aged samples (PA with $0.8$ aging steps per site) show no discernible size-dependence, with increasing amount of aging a rather weak system size dependence emerges, particularly near the stress peak, as shown in the inset.  
\begin{figure*}[t!]
\includegraphics[width = \textwidth]{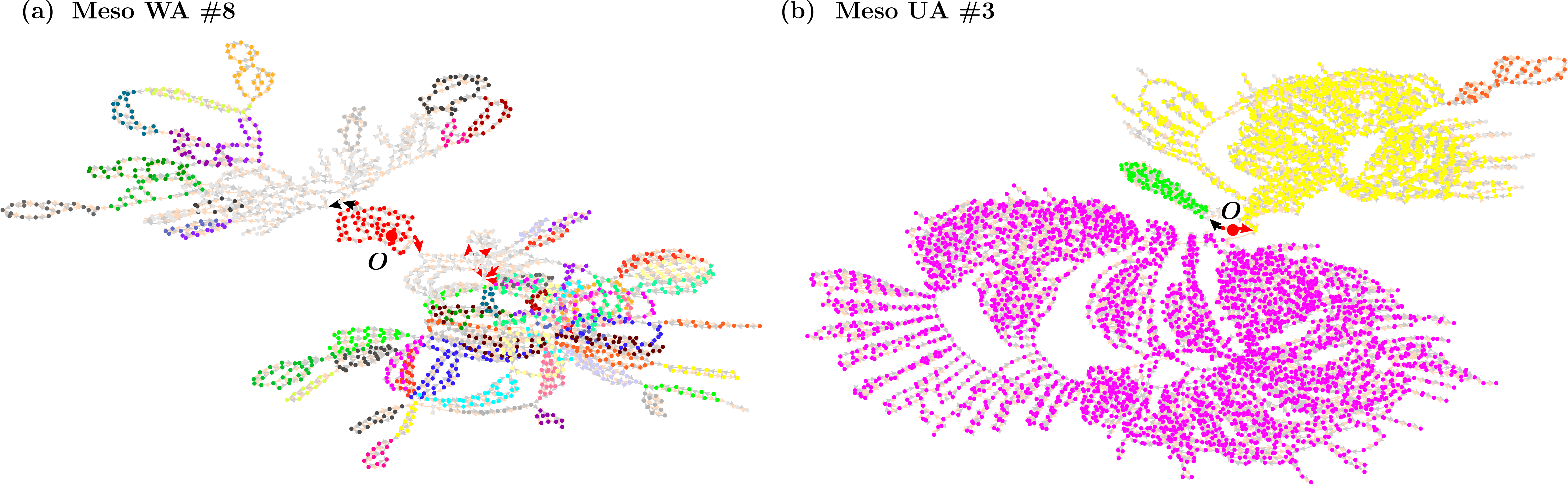}
\caption{Excerpts of transition graphs extracted from well-aged (WA) and ultra-well-aged (UA) mesoscopic glasses with $N = 32$. Refer to text  for further details.}
\label{fig:tgraphs_wa}
\end{figure*}

\section{Catalogs extracted from simulations of the atomistic model and the mesoscopic model with well-aged reference configurations. \label{sec:catalog-properties}}

In addition to the Tables \ref{tab:N32PAcatalogs} and \ref{tab:N32WAcatalogs} in the main text, which describe the properties of catalogs extracted from moderately- and poorly-aged glasses of our mesoscopic model with $N=32$, we also list here the properties of (i) two increasingly better-aged mesoscopic catalogs, prepared from glasses subjected to $150$ and $4000$ aging steps per site, which we will refer to as the well-aged (WA) and ultra-aged (UA) glasses, respectively, and (ii) atomistic catalogs obtained from $8$ poorly-aged and $30$ well-aged reference configurations. The aging of the atomistic glasses is controlled by the rate of quenching to zero temperature from a high temperature liquid, as described in Appendix \ref{sec:atomistic-simulations}. We refer to these as fast (AS Fast) and slowly (AS Slow) quenched atomistic glasses, respectively. 

\begin{table}[h!]
\caption{Properties of the $10$ catalogs obtained from the well-aged (WA) reference configurations aged at $150$ aging steps per site of our mesoscopic model. Refer to the caption of Table \ref{tab:N32PAcatalogs} for the description of the columns.}
\rowcolors{2}{white}{black!20}
\begin{tabular}{ *8l }  \toprule \hline
Run  &  $g_{\rm comp}$  & $N_0$ & $N_{\rm SCC}$ & $n_{\rm cycles}$ & $n_{\rm supp SCC}$ & $s^{\rm max}_{\rm supp SCC}$ & $n_{\rm cycles}^{\rm max suppSCC}$ \\ \midrule \hline 
1 & 39 & 79565 & 25151 & 130 & 44 & 60 & 50 \\ 
2 & 39 & 91201 & 27337 & 60 & 33 & 475 & 1 \\ 
3 & 39 & 114686 & 36931 & 305 & 14 & 300 & 219 \\ 
4 & 39 & 124298 & 33459 & 525 & 107 & 344 & 32 \\ 
5 & 39 & 38629 & 13207 & 685 & 181 & 419 & 12 \\ 
6 & 39 & 64475 & 13240 & 1388 & 606 & 127 & 7 \\ 
7 & 39 & 26421 & 6677 & 115 & 45 & 38 & 12 \\ 
8 & 39 & 80317 & 37092 & 122 & 65 & 56 & 50 \\ 
9 & 39 & 68154 & 17009 & 43 & 5 & 26 & 38 \\ 
10 & 39 & 84064 & 34515 & 66 & 2 & 88 & 42 \\ 
 \midrule \hline 
{\bf ALL} &  n/a  & 771810 & 244618 & 3439 & 1102 &  n/a  & 463 \\ \bottomrule
\hline
\end{tabular}
\end{table}

\begin{table}[h!]
\caption{Properties of the $10$ catalogs obtained from a ultra-aged (UA) reference configurations aged at $4000$ aging steps per site of our mesoscopic model. Refer to the caption of Table \ref{tab:N32PAcatalogs} for the description of the columns.}
\rowcolors{2}{white}{black!20}
\begin{tabular}{ *8l }  \toprule \hline
Run  &  $g_{\rm comp}$  & $N_0$ & $N_{\rm SCC}$ & $n_{\rm cycles}$ & $n_{\rm supp SCC}$ & $s^{\rm max}_{\rm supp SCC}$ & $n_{\rm cycles}^{\rm max suppSCC}$ \\ \midrule \hline 
1 & 45 & 24999 & 2714 & 162 & 57 & 433 & 1 \\ 
2 & 45 & 22443 & 1758 & 486 & 114 & 625 & 72 \\ 
3 & 45 & 25541 & 1834 & 468 & 79 & 3173 & 43 \\ 
4 & 45 & 28065 & 5796 & 205 & 77 & 703 & 5 \\ 
5 & 45 & 77224 & 24002 & 94 & 51 & 168 & 8 \\ 
6 & 45 & 19225 & 1643 & 314 & 107 & 2489 & 4 \\ 
7 & 45 & 17750 & 1394 & 300 & 104 & 1292 & 1 \\ 
8 & 45 & 15036 & 1066 & 681 & 107 & 1428 & 35 \\ 
9 & 45 & 68780 & 14479 & 94 & 54 & 60 & 2 \\ 
10 & 45 & 17118 & 1129 & 911 & 161 & 1467 & 93 \\ 
 \midrule \hline 
{\bf ALL} &  n/a  & 316181 & 55815 & 3715 & 911 &  n/a  & 264 \\ \bottomrule
\hline
\end{tabular}
\end{table}

Fig.~\ref{fig:tgraphs_wa} shows an excerpt of the transition graph extracted from samples of sample $\#8$ of our WA glass, panel (a), and sample $\#3$ of the ultra-stable UA glass, panel (b). The number of mesostates displayed in the $t$-graph excerpts shown are  $1665$ and $4610$, respectively.
We have obtained the graphs shown in Fig.~\ref{fig:tgraphs} of the main text as well as in Fig.~\ref{fig:tgraphs_wa} by starting out in the reference configuration and following SCCs and the transitions between them until at least $1500$ mesostates have been collected. For every SCC reached in this way, we added also the remaining mesostates belonging to that SCC so that the total number of vertices constituting the graph excerpt is typically larger than $1500$. The number of SCCs shown in the excerpts of the two graphs in Fig.~\ref{fig:tgraphs_wa} are $216$ (WA) and $19$ (UA). The sizes of SCCs seen in the WA excerpt  are small ($s_{\rm SCC} \le 56$), while the UA excerpt has three very large SCC with sizes $s_{\rm SCC} = 3173, 1271$, and $82$, shown in pink, yellow, and green, respectively. These findings are consistent with the SCC scatter plots shown in panels (e) and (f) of Fig.~\ref{fig:scc-scatter}. We believe that the emergence of the giant SCCs in the ultra-stable sample is a finite-size effect. 

The stress-strain curves of the well-aged samples under uniform shear exhibit large stress changes across the yielding transition. For the WA and UA samples shown in Fig.~\ref{fig:tgraphs_wa}, the magnitude of these stress-jumps under shear in the forward and reverse directions are $\Delta \sigma = 0.46, 0.45$ for the WA glass and $\Delta \sigma = 0.62, 0.70$ for the ultra-stable UA glass. In the graphs shown in Fig.~\ref{fig:tgraphs_wa} we have highlighted  transitions that involve stress-jumps with a magnitude of at least $0.1$, by fat black ($\Up$-transition) and red arrows ($\Dn$-transition). Despite of the relative low threshold value chosen for these jumps, only very few transitions in the two graphs shown experience large stress changes. Note that for both the WA and UA samples the transitions involving the large stress-jumps under forward and reverse shear tend to partition the graph into two halves (at least to the resolution of the number of vertices shown). This effect is even more dramatic for the ultra-stable glass sample where the transitions with large stress jumps immediately leads to giant SCCs.

\begin{table}[h!]
\caption{Properties of the $8$ catalogs obtained from poorly-aged (fast quench) reference configurations of our atomistic model. Refer to the caption of Table \ref{tab:N32PAcatalogs} for the description of the columns.}
\rowcolors{2}{white}{black!20}
\begin{tabular}{ *8l }  \toprule \hline
Run  &  $g_{\rm comp}$  & $N_{0}$ & $N_{\rm SCC}$ & $n_{\rm cycles}$ & $N^{\rm supp}_{\rm SCC}$ & $s^{\rm max}_{\rm supp SCC}$ & $n_{\rm cycles}^{\rm max SCC}$ \\ \midrule \hline 
1 & 40 & 57638 & 24123 & 4650 & 1617 & 929 & 215 \\ 
2 & 43 & 56158 & 27733 & 4515 & 1451 & 413 & 217 \\ 
3 & 37 & 55658 & 24119 & 5380 & 1305 & 106 & 9 \\ 
4 & 36 & 55057 & 24931 & 6972 & 1901 & 244 & 255 \\ 
5 & 41 & 57602 & 27645 & 3297 & 834 & 458 & 396 \\ 
6 & 35 & 53114 & 27939 & 4694 & 1453 & 259 & 244 \\ 
7 & 41 & 65842 & 29580 & 4323 & 1185 & 379 & 253 \\ 
8 & 45 & 58439 & 24794 & 5068 & 1187 & 234 & 235 \\ 
 \midrule \hline 
{\bf ALL} &  n/a  & 459508 & 210864 & 38899 & 10933 &  n/a  & 1824 \\ \bottomrule
\hline
\end{tabular}
\label{tab:glv2catalogs}
\end{table}

\begin{table}[h!]
\caption{Properties of the $30$ catalogs obtained from well-aged (slow quench) references configurations of  our atomistic model. Refer to the caption of Table \ref{tab:N32PAcatalogs} for the description of the columns.}
\rowcolors{2}{white}{black!20}
\begin{tabular}{ *8l }  \toprule \hline
Run  &  $g_{\rm comp}$  & $N_{0}$ & $N_{\rm SCC}$ & $n_{\rm cycles}$ & $N^{\rm supp}_{\rm SCC}$ & $s^{\rm max}_{\rm supp SCC}$ & $n_{\rm cycles}^{\rm max SCC}$ \\ \midrule \hline 
1 & 37 & 21105 & 11892 & 908 & 118 & 642 & 285 \\ 
2 & 39 & 16416 & 7202 & 797 & 211 & 196 & 148 \\ 
3 & 34 & 13894 & 5774 & 1101 & 264 & 503 & 135 \\ 
4 & 34 & 13710 & 5188 & 874 & 246 & 328 & 153 \\ 
5 & 37 & 18417 & 10118 & 552 & 140 & 371 & 120 \\ 
6 & 40 & 17618 & 6659 & 1660 & 275 & 718 & 302 \\ 
7 & 31 & 21250 & 10511 & 1373 & 371 & 330 & 64 \\ 
8 & 37 & 20940 & 8876 & 907 & 349 & 802 & 170 \\ 
9 & 43 & 18145 & 8578 & 664 & 26 & 980 & 78 \\ 
10 & 34 & 16847 & 6895 & 793 & 200 & 334 & 33 \\ 
11 & 32 & 13849 & 7535 & 766 & 201 & 521 & 178 \\ 
12 & 40 & 17723 & 7326 & 1200 & 233 & 356 & 148 \\ 
13 & 36 & 19814 & 10441 & 465 & 106 & 343 & 193 \\ 
14 & 41 & 22248 & 8824 & 1132 & 220 & 513 & 251 \\ 
15 & 35 & 14288 & 7221 & 727 & 123 & 377 & 199 \\ 
16 & 39 & 20930 & 7124 & 1562 & 257 & 2688 & 712 \\ 
17 & 42 & 15207 & 4769 & 1884 & 252 & 1017 & 375 \\ 
18 & 42 & 20779 & 7891 & 1719 & 271 & 1232 & 583 \\ 
19 & 35 & 16019 & 7276 & 712 & 186 & 729 & 263 \\ 
20 & 39 & 17477 & 6786 & 765 & 169 & 1489 & 307 \\ 
21 & 37 & 22784 & 11117 & 447 & 78 & 486 & 226 \\ 
22 & 38 & 17773 & 6717 & 473 & 96 & 853 & 181 \\ 
23 & 34 & 18273 & 7791 & 549 & 180 & 118 & 48 \\ 
24 & 42 & 24157 & 8904 & 1099 & 256 & 451 & 102 \\ 
25 & 37 & 14743 & 4966 & 1092 & 207 & 1199 & 274 \\ 
26 & 39 & 19295 & 9505 & 664 & 192 & 383 & 89 \\ 
27 & 36 & 19070 & 9547 & 873 & 301 & 518 & 146 \\ 
28 & 39 & 23452 & 10941 & 786 & 105 & 892 & 172 \\ 
29 & 42 & 18683 & 8289 & 822 & 89 & 833 & 341 \\ 
30 & 39 & 20426 & 9671 & 1493 & 141 & 963 & 883 \\ 
 \midrule \hline 
{\bf ALL} &  n/a  & 555332 & 244334 & 28859 & 5863 &  n/a  & 7159 \\ \bottomrule
\hline
\end{tabular}
\label{tab:waglv2catalogs}
\end{table}

\bibliography{bib_cycling.bib}

\end{document}